\documentclass[a4paper,12pt]{article}
\usepackage{amsmath,amssymb}
\usepackage{pstricks}
\usepackage{graphicx}
\usepackage{inputenc}
\newpsobject{grilla}{psgrid}{subgriddiv=1,
  griddots=10,gridlabels=6pt}
\newcommand{\AddrAHEP}{
  {\it AHEP Group, Instituto de F\'{\i}sica Corpuscular --
  C.S.I.C./Universitat de Val{\`e}ncia \\
  Edificio de Institutos de Paterna, Apartado 22085,
  E--46071 Val{\`e}ncia, Spain}}

\begin{document}

\begin{titlepage}
  \begin{flushright}
    hep-ph/0604012 \\
    IFIC/06-08\\
  \end{flushright}
  \vspace*{3mm}  
  
  \begin{center}
    \textbf{{\large Leptonic Charged Higgs Decays  
        in the Zee Model}}\\[10mm]
    {D. Aristizabal Sierra$^{1}$ and Diego Restrepo$^{2}$}\vspace{0.3cm}\\
    $^{1}$\AddrAHEP.\vspace{0.3cm}\\
    $^{2}${\it Instituto de F{\'\i}sica,  
    Universidad de Antioquia \\
    A.A. 1226 \ Medell{\'\i}n,  \ Colombia.
    \vspace{0.7cm}}
  \end{center}

  \begin{abstract}
    \noindent
    \baselineskip=23pt
    We consider the version of the Zee model where both Higgs 
    doublets couple to leptons. Within this framework 
    we study charged Higgs decays. We focus on a model with 
    minimal number of parameters consistent with 
    experimental neutrino data. Using constraints from 
    neutrino physics we
    (i) discuss the reconstruction of the parameter space
    of the model using the leptonic decay patterns of
    both of the two charged Higgses, 
    $h_{1,2}^{+}\rightarrow \ell_{j}^{+}\nu_{i}$, and the
    decay of the heavier charged Higgs, 
    $h_{2}^{+}\rightarrow h^{+}_{1}h^{0}$;
    (ii) show that the decay rate 
    $\Gamma(h_{1}^{+}\rightarrow \mu^{+}\nu_{i})$ in general
    is enhanced in comparision to the standard two Higgs 
    doublet model while in some regions of parameter space
    $\Gamma(h_{1}^{+}\rightarrow \mu^{+}\nu_{i})$ 
    even dominates over
    $\Gamma(h_{1}^{+}\rightarrow \tau^{+}\nu_{i})$.
  \end{abstract}
\end{titlepage}
\section{Introduction}
Neutrino oscillation experiments, including the results of
KamLAND \cite{Eguchi:2002dm} have confirmed the LMA-MSW 
oscillation solution of the solar neutrino problem. 
Together with the earlier discoveries in atmospheric neutrinos 
\cite{Fukuda:2001nj}, one can be fairly confident that all 
neutrino flavours mix and that at least two non-zero neutrino 
masses exist. 

In the standard model neutrinos are massless. 
Among all the existing models to generate small neutrino Majorana 
masses the seesaw mechanism \cite{Gell-Mann:vs}
is perhaps the most popular. However, 
this is not the only theoretical approach 
to neutrino masses. Other possibilities include
Higgs triplets \cite{Schechter:1980gr}, 
supersymmetric models with broken $R$-parity 
\cite{hall:1984id,Hirsch:2000ef}, some hybrid mechanisms 
that combine the triplet and the $R$-parity ideas 
\cite{AristizabalSierra:2003ix} and radiative 
mechanisms \cite{Zee:1980ai,Babu:1988ki}.

Here we consider a particular radiative mechanism,  
the Zee model \cite{Zee:1980ai}. In this model the scalar sector 
of the standard model is enlarged to include a charged SU(2)
gauge singlet scalar and a second Higgs doublet. This particle
content allows to write an explicit lepton number ($L$) violating 
term in the scalar potential and leads to neutrino masses
at one loop order. In the Minimal Zee Model
(MZM), only one Higgs doublet couples to 
leptons \cite{Wolfenstein:1980sy}. As a result, dangerous 
Flavour Changing Neutral Current (FCNC)
processes are forbidden.
It has been shown \cite{He:2003ih} that combining SNO, 
KamLAND and K2K experimental data this version is ruled 
out.

However, this does not mean that the Zee model is ruled out. The 
original version, from now on called the 
General Zee Model (GZM) \cite{Balaji:2001ex},
in which both of the two Higgs doublets 
couple to the matter fields has been shown 
\cite{He:2003ih,Balaji:2001ex,Hasegawa:2003by}
to be consistent with atmospheric and solar neutrino 
data as well \cite{Maltoni:2004ei}. 

Once one allows both of the Higgs doublets to couple to leptons 
the number of model parameters increases. Here instead 
of working with all the couplings of the model we will 
consider a scheme, previously discussed in references 
\cite{Balaji:2001ex,Hasegawa:2003by},
where the neutrino mass matrix has a two-zero-texture.
This particular GZM will be called Next to MZM (NMZM). 

In the Higgs sector, after spontaneous breaking of the
electroweak symmetry, the charged gauge singlet mixes with
the charged components of the two Higgs doublets.
The resulting charged Higgs eigenstates 
($h_{i}^{\pm}$ with $i=1,2$) decay to states with 
charged leptons and neutrinos. These decays can be used, 
in principle, to reconstruct the Majorana neutrino mass matrix.

We will show that due to the constraints imposed by 
neutrino physics, the $Br(h^{+}_{1}\to\sum_{i}\nu_{i}\mu^{+})$ 
is enhanced in comparision to the two-Higgs 
doublet models (2HDM) of type-I and type-II 
\footnote{In type-I only one of the
Higgs fields couples to the SM fermions, in type-II one Higgs
field couples to up-type quarks and the other Higgs field
couples to down-type quarks. There is another version called 
type-III \cite{Hou:1991un} where both Higgs fields couple 
to all SM fermions.}. 
Moreover, we will show that in large parts of the parameter 
space $Br(h^{+}_{1}\to\sum_{i}\nu_{i}\mu^{+})\gtrsim
Br(h^{+}_{1}\to\sum_{i}\nu_{i}\tau^{+})$.
For details see section~\ref{sec:NMZMandCP}.

The rest of this paper is organized as follows. 
In section~\ref{sec:model} we give the generalities
of the GZM and work out the Higgs mass spectrum of the model.
In section~\ref{sec:chargedphen} we study
charged Higgs production at a future $e^{+}e^{-}$
collider. In section~\ref{sec:fcnc} we discuss
the bounds on the parameters of the model coming from FCNC
processes constraints. In section~\ref{sec:neutrinophysics} 
we describe the Majorana neutrino mass matrix 
within the GZM and in the NMZM. In 
section~\ref{sec:NMZMandCP} we discuss the connection 
between neutrino physics and charged Higgs decays. 
In section~\ref{sec:conclusions}
we present our conclusions and summarize our results.
\section{The Model}
\label{sec:model}
\subsection{Generalities}
\label{sec:generalities}
If no new fermions are added to the standard model 
neutrino masses must be always 
of Majorana type, i.e. the mass term must violate
$L$. In the Zee model an $L=2$ charged scalar, 
$h^{+}$, is introduced. Since this field carries electric
charge its vacumm expectation value (vev) must vanish.
Therefore in this model $L$ cannot be spontaneously broken.
However, $h^{+}$ can be used to drive the lepton number
breaking from the leptonic sector to the scalar sector. In order to
accomplish this a new SU(2)$_{L}$ doublet has to be added, 
as a result an explicit $L$ violation term can be
written. This term is given by
\begin{equation}
  \label{eq:leptonnumberviolationterm}
  \mu\epsilon_{\alpha\beta}H_{1}^{\alpha}
  H_{2}^{\beta}h^{-} + \mbox{H.c.}
\end{equation}
where $\mu$ is a coupling with dimension of mass and 
$H_{1}$ and $H_{2}$ are doublets with hypercharge
$Y_1=Y_2=1$.

The most general Yukawa couplings of the model can 
be written as
\begin{equation}
  \label{eq:zeeLag}
  -{\cal L}_{Y} = 
  \bar{L}_{i}(\Pi_{a})_{ij}H_{a}e_{Rj} + 
  \epsilon_{\alpha\beta}\bar{L}^{\alpha}_{i}
  f_{ij}\,C(\bar{L}^{T})^{\beta}_{j}h^{-} +
  \text{H.c.}\,,
\end{equation}
where $L_{i}$ are lepton doublets, $e_{Rj}$ are lepton singlets,
$C$ is the charge conjugation operator, $\Pi_{a}$ 
($a=1,2$) and $f$ are $3\times 3$
matrices in flavour space, $\epsilon_{\alpha\beta}$ 
($\alpha,\beta=1,2$) is the SU(2)$_{L}$ antisymmetric tensor 
and $i,j=1,2,3$ are family indices. $f$ is an antisymmetric 
matrix due to Fermi statistics.

In general both of the two Higgs doublets can acquire
vev's, $\langle H_a\rangle=v_a$, with $v=\sqrt{v_1^2+v_2^2}
\simeq 246$ GeV. As usual, the ratio of these vev's
can be parametrized as $\tan\beta=v_2/v_1$.
\subsection{Higgs potential and scalar mass spectrum}
\label{sec:higgspotential}
Though in this work we are interested mainly in the 
charged Higgs sector of the model and its relation
with neutrino physics, we will briefly discuss the 
full scalar mass spectrum.

The Higgs potential is invariant under a global SO(2) 
transformation
\begin{equation}
  \label{eq:so2globaltrans}
  \begin{pmatrix}
    H'_1\\
    H'_2
  \end{pmatrix}
  =
  \begin{pmatrix}
    \cos\beta   & \sin\beta\\
    - \sin\beta & \cos\beta 
  \end{pmatrix}
  \begin{pmatrix}
    H_1\\
    H_2
  \end{pmatrix}.
\end{equation}
Moreover, the Yukawa Lagrangian given in Eq.~(\ref{eq:zeeLag}) 
is also invariant under the above transformation if the
Yukawa matrices are appropriately rotated, namely
\begin{equation}
  \label{eq:matrixtransformations}
  \begin{pmatrix}
    \Pi_1'\\
    \Pi_2'
  \end{pmatrix}
  =
  \begin{pmatrix}
    \cos\beta   & \sin\beta\\
    - \sin\beta & \cos\beta 
  \end{pmatrix}
  \begin{pmatrix}
    \Pi_1\\
    \Pi_2
  \end{pmatrix}.
\end{equation} 
Therefore the full model is invariant under global
SO(2) transformations of the two Higgs doublets.
Thus, we are free to redefine our two doublet scalar 
fields by making an arbitrary SO(2)
transformation. A particular choice of fields corresponds
to a choice of basis. There is a basis in which only one
of the two Higgs doublets acquire a vev. In the context
of the 2HDM of type-III it is called the {\it Higgs basis}.
Notice that in this basis $\tan\beta=0$.

In the Higgs basis the most general gauge invariant scalar 
potential of the model, consistent with 
renormalizability reads
\begin{eqnarray}
  \label{eq:scalarpotentialinhiggsbas}
  V & = & \mu^{2}_{1}H_{1}^{\dagger}H_{1}
  + \mu^{2}_{2}H_{2}^{\dagger}H_{2}
  - [\mu^{2}_{3}H_{1}^{\dagger}H_{2} + \mbox{H.c.}]
  + \frac{1}{2}\lambda_{1}(H_{1}^{\dagger}H_{1})^{2}
  \nonumber\\
  && + \frac{1}{2}\lambda_{2}(H_{2}^{\dagger}H_{2})^{2}
  + \lambda_{3}(H_{1}^{\dagger}
  H_{1})(H_{2}^{\dagger}H_{2})
  + \lambda_{4}(H_{1}^{\dagger}
  H_{2})(H_{2}^{\dagger}H_{1})
  \nonumber\\
  && + \left\{
    \frac{1}{2}\lambda_{5}(H_{1}^{\dagger}H_{2})^{2}
    + [\lambda_{6}(H_{1}^{\dagger}H_{1})
    + \lambda_{7}(H_{2}^{\dagger}H_{2})]
    H_{1}^{\dagger}H_{2}
    +\mbox{H.c.}
  \right\}
  \nonumber\\
  && + \mu_{h}^{2}|h^{+}|^{2} + \lambda_{h}|h^{+}|^{4}
  + \lambda_{8}|h^{+}|^{2}H_{1}^{\dagger}H_{1} 
  + \lambda_{9}|h^{+}|^{2}H_{2}^{\dagger}H_{2}
  \nonumber\\
  && + \lambda_{10}|h^{+}|^{2}(H_{1}^\dagger H_{2}
  + \mbox{H.c.})
  + \mu\epsilon_{\alpha\beta}H_{1}^{\alpha}
  H_{2}^{\beta}h^{-}.
\end{eqnarray}
Since we will not deal with CP-violating effects
we only consider real coefficients

Minimization of the scalar potential, 
Eq.~(\ref{eq:scalarpotentialinhiggsbas}),
leads to the conditions \cite{Davidson:2005cw}
\begin{eqnarray}
  \label{eq:minimizationconditions}
  \mu_{1}^{2} &=& -\frac{1}{2}\lambda_{1}v^{2}
  \nonumber\\
  \mu_{3}^{2} &=& \frac{1}{2}\lambda_{6}v^{2}.
\end{eqnarray}
These conditions can be used to eliminate $\mu_{1}^{2}$ 
and $\mu_{3}^{2}$ as independent variables from $V$.

Of the original ten scalar degrees of freedom, three Goldstone 
bosons ($G^{\pm}$ and $G^{0}$) are absorbed by the $W^{\pm}$ and 
$Z^{0}$. The remaining seven physical Higgs particles are:
two CP-even ($h^{0}$ and $H^{0}$ with 
$m_{h^{0}}\leq m_{H^{0}}$), one CP-odd ($A^{0}$) and two charged 
Higgs pairs ($h_{1}^{\pm}$ and $h_{2}^{\pm}$).

In the basis $\mathbf{\Phi}^{\dagger}=(G^{-}, 
H^{-}, h^{-})$ 
the squared-mass matrix for the charged Higgs states is given by
\begin{equation}
  \label{eq:scalarM}
  {\cal M}_{C}^{2}=
  \begin{pmatrix}
    0   &          0         &  0                   \\
    0   &  M_{H^{\pm}}^{2}   &  -\mu v/\sqrt{2}     \\
    0       &  -\mu v/\sqrt{2} &  {\cal M}_{33}^{2}   
  \end{pmatrix}\;,
\end{equation}
where 
\begin{eqnarray}
  \label{eq:entriesofmassmatrix}
  M_{H^{\pm}}^{2} &=& \mu_{2}^{2} +\frac{1}{2}v^{2}\lambda_{3}
  \nonumber\\
  {\cal M}_{33}^{2} &=& \mu_{h}^{2} + v^{2}\lambda_{8}.
\end{eqnarray}
The matrix element $M_{H^{\pm}}^{2}$
corresponds to the squared-mass of the charged scalars
($H^{\pm}$) that in the absence of the SU(2)$_{L}$ 
singlets $h^{\pm}$ would be physical Higgs particles. 

The squared-mass matrix 
${\cal M}_{C}^{2}$ can be diagonalized
by the rotation matrix
\begin{equation}
  \label{eq:rotationM}
  R = 
  \begin{pmatrix}
    1 &      0       &     0       \\
    0 & \cos\varphi  &  \sin\varphi\\
    0 & -\sin\varphi &  \cos\varphi
  \end{pmatrix}\,.
\end{equation}
where the angle $\varphi$ characterize the size 
of the $H^{\pm}-h^{\pm}$ mixing. 

The mass eigenstate basis in the charged Higgs sector 
is defined as 
$\mathbf{H}^{\dagger}=(G^{-}, h^{-}_{1},h^{-}_{2})$
and the rotation angle is given by
\begin{equation}
  \label{eq:rotAforScalars}
  \sin2\varphi=\frac{\sqrt{2}v\mu}{M_{2}^2 - M_{1}^2}.
\end{equation}
Here $M_{1}$ and $M_{2}$ stand for the masses of the scalars 
$h_{1}^{\pm}$ and $h_{2}^{\pm}$ which are given by
\begin{equation}
  \label{eq:eigenvalues}
  M_{1,2}^2 = \frac{1}{2}
  \left(
  M_{H^{\pm}}^{2} + {\cal M}_{33}^{2} \mp
  \sqrt{(M_{H^{\pm}}^{2} - {\cal M}_{33}^{2})^{2}
  + 2\mu^{2}v^{2}}
  \right).
\end{equation}

In the Higgs basis the squared-masses for the CP-odd
and CP-even Higgs states are given by \cite{Davidson:2005cw}
\begin{align}
  \label{eq:cpevencpoddmasses}
  M_{A^{0}}^{2} &= M_{H^{\pm}}^{2} -\frac{1}{2}v^{2}
  (\lambda_5 - \lambda_4)
  \nonumber\\
  M_{H^{0},h^{0}}^{2} &= \frac{1}{2}
  \left[
    M_{A^{0}}^{2} + v^{2}(\lambda_1 + \lambda_5) \pm 
    \sqrt{[M_{A^{0}}^{2} + v^{2}(\lambda_5 - \lambda_1)]^2 
      + 4v^{4}\lambda_6^{2}}
  \right]
\end{align}
\section{Charged Scalar Phenomenology}
\label{sec:chargedphen}
\subsection{Cross section}
\label{sec:crosssec}
In the following we discuss charged scalar $h^{\pm}_{k}$
($k=1,2$) production at a future $e^{+}e^{-}$ collider. 
$h^{\pm}_{k}$ are produced in $e^{+}e^{-}$ 
annihilation via s-channel exchange of a $\gamma$ or $Z^{0}$
\footnote{There is also a t-channel Yukawa production 
through neutrino exchange but due to the smallness of 
this contribution to the total production cross section 
we do not consider it here.}. 
The total cross section
for the process $e^{+}e^{-}\to h_{k}^{+}h_{k}^{-}$ 
will be the sum of three terms
\begin{figure}[t]
  \centering
  \begin{pspicture}(-4,-0.7)(4,7)
    \uput[r](-4,3)
    {\mbox{\includegraphics[width=9cm,height=7.5cm]
        {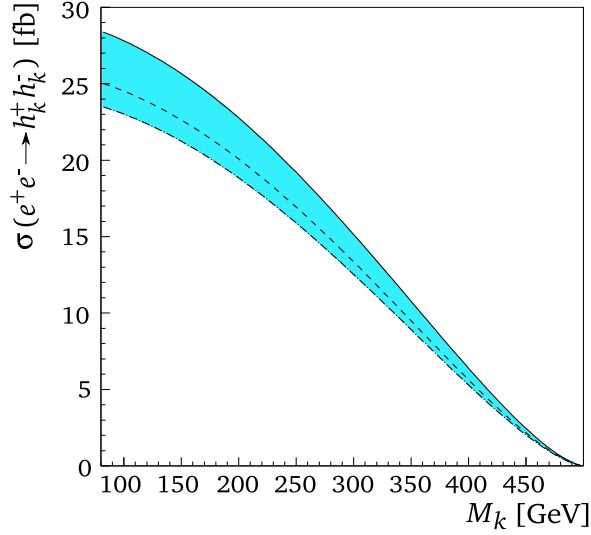}}}
  \end{pspicture}
  \caption{Production cross section for charged 
    scalars $h_{k}^{\pm}$ at an 1 TeV $e^{+}e^{-}$ 
    collider with unpolarized beams.}
  \label{fig:seficaz}
\end{figure}
\begin{equation}
  \label{eq:crossSection}
  \sigma_{\mbox{\tiny{total}}} = \sigma_{\gamma} + \sigma_{Z} + 
  \sigma_{\gamma Z}
\end{equation}
corresponding to the pure photon, pure $Z$ and 
photon--$Z$ interference contributions
respectively. Thus
\begin{eqnarray}
  \label{eq:crosssectionbypieces}
  \sigma_{\gamma}& = &\frac{1}{48\pi}\beta^{3}
  (g\,s_{w})^{4}\frac{1}{s}
  \\
  \sigma_{Z}& = &\frac{1}{3072\pi}\beta^{3}
  \frac{g^{4}}{c^{4}_{w}}(W_{1k}^{2}-2s_{w}^{2})^{2}
  [(-1+4s_{w}^{2})^{2}+1]
  \nonumber\\
  &&
  \frac{s}{(s-M_{Z}^{2})^{2}+M_{Z}^{2}\Gamma_{Z}^{2}}\\
  \sigma_{Z\gamma}& = &-\frac{1}{192\pi}\beta^{3}
  \frac{g^{4}s_{w}^{2}}{c_{w}^{2}}
  (W_{1k}^{2}-2s_{w}^{2})(-1+4s_{w}^{2})
  \nonumber\\
  &&
  \frac{s-M^{2}_{Z}}{(s-M_{Z}^{2})^{2} + 
    M_{Z}^{2}\Gamma_{Z}^{2}}.
\end{eqnarray} 
where $s_{w}=\sin\theta_{w}$, $c_{w}=\cos\theta_{w}$,
\begin{equation}
  \label{eq:beta}
  \beta=\sqrt{1-4\frac{M_{k}^{2}}{s}}
\end{equation}
and $W_{11}=W_{22}=\cos\varphi$ and 
$W_{12}=-W_{21}=\sin\varphi$.

From Eqs. (\ref{eq:entriesofmassmatrix}), 
(\ref{eq:eigenvalues}) and (\ref{eq:cpevencpoddmasses}) 
it can be noted that fixing $M_{1,2}$ does not fix 
$M_{H^{0},h^{0}}$ and $M_{A^{0}}$. 
Therefore, it is possible to take $M_{1,2}$ and $W_{1k}$
as free parameters without being in conflict with LEP
bounds for the CP-even and CP-odd Higgs masses
\cite{Heister:2002ev,Abbiendi:2000ug,Abdallah:2004wy}.

In Fig.~\ref{fig:seficaz} we show the cross section at an
1 TeV $e^{+}e^{-}$ collider with unpolarized beams.
There we have taken $80\,\text{Gev}\leq M_{k}\leq 500\,\text{GeV}$.
The spread in the plot is due to the dependence of the cross sections
on $\varphi$. It is important to notice that for small (large) values
of $\varphi$ the cross section for $h^{+}_{1}$ increases (decreases)
while the cross section for $h^{+}_{2}$ decreases (increases).
Figure~\ref{fig:seficaz} illustrates the situation for the case
$\varphi=0$. In that case $h^{+}_{1}$ coincides with the SU(2)
doublet $H^{+}$ (solid line) and $h^{+}_{2}$ with the 
SU(2) singlet $h^{+}$ (dashed line). The dotted-dashed line
corresponds to $\cos\varphi\simeq 0.6$.

In Fig.~\ref{fig:seficaz} it can be seen that up to a mass 
of $\sim350$ GeV the charged scalars have a cross section 
larger than 10 fb. Assuming an integrated luminosity of 
1 ab$^{-1}$ this implies that at least 10$^4$ charged scalar 
pairs will be produced.
\subsection{Decay Widths}
\label{sec:decaywidth}
After the spontaneous electroweak symmetry breaking charged leptons
acquire mass, namely
\begin{equation}
  \label{eq:chargedLMM}
  \widehat{M}_{\ell} = \frac{1}{\sqrt{2}}\sum_{a}v_{a}\Pi_{a}.
\end{equation}

In the mass eigenstate basis for the charged scalars
we have
\begin{eqnarray}
  \label{eq:lagrangianinME}
  -{\cal L}_{Y}& \supset &\bar{\nu}_{Li'}O_{i'j}
  e_{Rj}(\cos\varphi h^{+}_{1} - \sin\varphi h^{+}_{2})
  \nonumber\\
  &&+ (\nu_{Li})^{T}C(2f_{ij})e_{Lj}
  (\sin\varphi h^{+}_{1} + 
  \cos\varphi h^{+}_{2}) + 
  \text{H.c.}
\end{eqnarray}
where, in general, the couplings $O_{ij}$
are given by
\begin{equation}
  \label{eq:oijcoupling}
  O =
    -\sqrt{2}\frac{\tan\beta}{v}\widehat{M}_{\ell}
    +
    \frac{1}{\cos\beta}\Pi_{2}.
\end{equation}

Charged scalars $h_{1,2}^{+}$ will decay through the couplings
$O_{ij}$  and $f_{ij}$. Possible leptonic final 
states are $\nu_{i}\ell_{j}^{+}$. Possible final states involving
quarks are $\bar{d}_{i}u_{j}$. These decays are determined 
by the couplings $O^{q}_{ij}$ where, in general
\begin{equation}
  \label{eq:quarkcouplins}
  O^{q} =
    -\sqrt{2}\frac{\tan\beta}{v}\widehat{M}_{q}
    +
    \frac{1}{\cos\beta}\Pi_{2}^{q}.
\end{equation}
Here $q$ refers to up-type and down-type quarks, 
$\widehat{M}_{q}$ are the diagonal quark 
mass matrices, and $\Pi_{2}^{q}$ are $3\times3$ Yukawa 
coupling matrices of the second Higgs doublet. Notice that
in the Higgs basis $\widehat{M}_{\ell} = (v/\sqrt{2})\Pi_{1}$
and $O=\Pi_{2}$.

We are interested in the widths and branching ratios
for leptonic final states. The Lagrangian 
(\ref{eq:lagrangianinME}) determines the two body decays 
$h_{1,2}^{+}\to(\sum_{i}\nu_{i})\ell_{j}^{+}$. 
The decay rate reads
\begin{equation}
  \label{eq:decaywidths}
  \Gamma(h_{k}^{+}
  \to(\sum_{i}\nu_{i})\ell_{j}^{+})=
  \frac{M_{k}}{16\pi}\sum_{i}[O_{ij}^{2}W_{1k}^{2}
  + 
  (2f_{ij})^{2}W_{2k}^{2}]\,.
\end{equation} 
The couplings $h^{+}_{k}W^{-}Z$ and $h^{+}_{k}W^{-}\gamma$
do not exist in the Zee model. This can be understood
as follows: since $h^{+}_{k}$ is a mixture of $H^+$
and $h^+$ these couplings are determined by the SU(2)
doublet component. However, in the 2HDM of type-III these 
vertices do not exist \cite{Grifols:1980uq}. 
Therefore the decays $h_{k}^{+}\rightarrow W^{+}\gamma$, $W^{+}Z^{0}$ in
the Zee model are not present at tree level. For this reason
we do not consider them.
\begin{table}[t]
\centering
  \begin{tabular}{|c||c|}\hline\hline
   Process & Constraint\\\hline
   $\mu^{-}\rightarrow e^{+}e^{-}e^{-}$&
   $|O_{12}O_{11}|<3.6\times10^{-7}
   \left(\frac{M_h^{0}}{100\text{GeV}}\right)^{2}$\\
   $\tau^{-}\rightarrow e^{+}e^{-}e^{-}$&
   $|O_{13}O_{11}|<1.3\times10^{-3}
   \left(\frac{M_h^{0}}{100\text{GeV}}\right)^{2}$\\
    $\tau^{-}\rightarrow \mu^{+}\mu^{-}\mu^{-}$&
   $|O_{13}O_{12}|<0.9\times10^{-3}
   \left(\frac{M_h^{0}}{100\text{GeV}}\right)^{2}$\\
   $\tau^{-}\rightarrow \mu^{-}\mu^{-}e^{+}$&
   $|O_{23}O_{21}|<0.9\times10^{-3}
   \left(\frac{M_h^{0}}{100\text{GeV}}\right)^{2}$\\
   $\tau^{-}\rightarrow e^{-}\mu^{-}e^{+}$&
   $|O_{13}O_{21}+O_{23}O_{11}|<1.0\times10^{-3}
   \left(\frac{M_h^{0}}{100\text{GeV}}\right)^{2}$\\
   $\tau^{-}\rightarrow e^{-}\mu^{-}\mu^{+}$&
   $|O_{13}O_{22}+O_{23}O_{12}|<1.0\times10^{-3}
   \left(\frac{M_h^{0}}{100\text{GeV}}\right)^{2}$\\
   $\mu^{-}\rightarrow e^{-}\gamma$&
   $|O_{12}O_{11}+O_{22}O_{21}+O_{32}O_{31}|<4.1\times10^{-5}
   \left(\frac{M_h^{0}}{100\text{GeV}}\right)^{2}$\\
   $\tau^{-}\rightarrow e^{-}\gamma$&
   $|O_{13}O_{11}+O_{23}O_{21}+O_{33}O_{31}|<4.7\times10^{-2}
   \left(\frac{M_h^{0}}{100\text{GeV}}\right)^{2}$\\
   $\tau^{-}\rightarrow \mu^{-}\gamma$&
   $|O_{13}O_{12}+O_{23}O_{22}+O_{33}O_{32}|<3.3\times10^{-2}
   \left(\frac{M_h^{0}}{100\text{GeV}}\right)^{2}$\\
   \hline\hline
  \end{tabular}
  \caption{Constraints on the parameters $O_{ij}$
   from tree level and radiative FCNC processes
   induced by the neutral Higgs $h^0$.}
  \label{tab:fcnch0}
\end{table}
\section{Constraints from FCNC processes}
\label{sec:fcnc}
In the GZM FCNC interactions are induced by the 
charged and neutral Higgses. Bounds on the 
$O_{ji}O_{km}$ couplings can be obtained from
the non-observation of tree-level processes 
$\ell_{i}^{-}\rightarrow\ell_{j}^{+}\ell_{k}^{-}\ell_{m}^{-}$. 
Constraints on $O_{ki}O_{kj}$ come from radiative processes 
$\ell_{i}^{-}\rightarrow\ell_{j}^{-}\gamma$ induced
by neutral Higgses. Limits on $f_{ik}f_{kj}$ and on
$O_{ki}f_{kj}$ couplings come from radiative processes 
mediated by charged scalars\footnote{These processes
also give bounds on $O_{ik}O_{kj}$. However they are
weaker than those coming from radiative processes mediated
by neutral Higgses.}. An important remark is that once the
constraints on the $f_{ik}f_{kj}$ and $O_{ki}O_{kj}$
couplings are satisfied the limits on $O_{ki}f_{kj}$ are 
no longer important, for this reason we do not
list them. Table~\ref{tab:fcnch0} shows
the constraints coming from the processes mediated by neutral
scalars. Table~\ref{tab:fcnchc} summarize the limits on the 
$f_{ij}$ parameters. Experimental constraints used in 
both tables were taken from \cite{Eidelman:2004wy}
\section{Neutrino Physics}
\label{sec:neutrinophysics}
\subsection{Neutrino Mass Matrix in the GZM}
\label{sec:nmmintheZeem}
In this section we will discuss the neutrino mass 
matrix. The Majorana neutrino mass matrix in the Zee 
model arises at the one loop level through the exchange of the 
scalars $h^{\pm}_{1}$ and $h^{\pm}_{2}$ as shown in 
Fig.~\ref{fig:loops}. Assuming $M_{1}, M_{2}\gg 
m_{e}, m_{\mu}, m_{\tau}$ we have
\begin{equation}
  \label{eq:massmatrix}
  (M_{\nu})_{ii'}=\kappa
  [f_{ij}(\widehat{M}_{\ell})_{jj}O_{i'j} + 
  O_{ij}(\widehat{M}_{\ell})_{jj}f_{i'j}]
\end{equation}
where
\begin{equation}
  \label{eq:kapa}
  \kappa = \frac{\sin2\varphi}{(4\pi)^{2}}
  \ln\left(\frac{M^{2}_{2}}{M^{2}_{1}}\right).
\end{equation}
\begin{table}[t]
\centering
  \begin{tabular}{|c||c|}\hline\hline
   Process & Constraint\\\hline
   $\mu^{-}\rightarrow e^{-}\gamma$&
   $|f_{23}f_{13}|<4.1\times10^{-5}
   \left(\frac{M_1}{100\text{GeV}}\right)^{2}$\\
   $\tau^{-}\rightarrow e^{-}\gamma$&
   $|f_{23}f_{12}|<4.7\times10^{-2}
   \left(\frac{M_1}{100\text{GeV}}\right)^{2}$\\
   $\tau^{-}\rightarrow \mu^{-}\gamma$&
   $|f_{13}f_{12}|<3.3\times10^{-2}
   \left(\frac{M_1}{100\text{GeV}}\right)^{2}$\\
   \hline\hline
  \end{tabular}
  \caption{Constraints on the parameters $f_{ij}$
   coming from radiative FCNC processes induced 
   by the charged Higgs $h_{1}^{\pm}$.}
  \label{tab:fcnchc}
\end{table}
\subsection{Neutrino Mass Matrix in the NMZM}
\label{sec:nmmintheNMZM}
In this section we discuss the neutrino mass matrix in the 
context of the NMZM. In our scheme the neutrino mass 
matrix is assumed to be
\begin{equation}
  \label{eq:nmzmNMM}
  M_{\nu}=\kappa
  \begin{pmatrix}
    M_{ee}     &   M_{e\mu}    & M_{e\tau}   \\
    M_{e\mu}   &      0        & M_{\mu\tau} \\
    M_{e\tau}  &   M_{\mu\tau} &    0
  \end{pmatrix}.
\end{equation}

The neutrino mass matrix, can be diagonalized by a matrix
$U$, which can be parametrized as
\begin{equation}
  \label{eq:MNSmatrix}
  U=
  \begin{pmatrix}
    1 &    0    & 0     \\
    0 & c_{23}  & s_{23} \\
    0 & -s_{23} & c_{23}
  \end{pmatrix}
  \times
  \begin{pmatrix}
    c_{13}  &    0    & s_{13} \\
      0     &   1     &   0    \\
    -s_{13} &   0     & c_{13}
  \end{pmatrix}
  \times
  \begin{pmatrix}
    c_{12}  &    s_{12}& 0 \\
    -s_{12} & c_{12}   & 0 \\
        0  &    0     & 1
  \end{pmatrix},
\end{equation}
where $s_{ij}=\sin\theta_{ij}$ and $c_{ij}=\cos\theta_{ij}$.
Phases are zero since only real parameters are considered.

From
\begin{equation}
  \label{eq:diagonalizationofNM}
  U^{T}M_{\nu}U=\widehat{M_{\nu}},
\end{equation}
and taking the limit $\sin^{2}\theta_{13}=0$,
since experimental neutrino data require $\sin^{2}\theta_{13}$
to be small \cite{Maltoni:2004ei},
we can find approximate analytical 
expressions for the atmospheric and solar mixing angle
as well as for $\Delta m_{23}^{2}$:
\begin{eqnarray}
  \label{eq:neutrinoexpquant}
  \tan^{2}\theta_{23} &\simeq&
  \left(\frac{M_{e\tau}}{M_{e\mu}}\right)^2,
  \nonumber\\
  \tan 2\theta_{12} &\simeq&
  \sqrt{2}\frac{M_{e\mu} - M_{e\tau}}
  {M_{ee} + M_{\mu\tau}},
  \\
  \sqrt{\Delta m_{23}^{2}} &\simeq& 
  \frac{\kappa}{\sqrt{2}}(M_{e\mu} - M_{e\tau}).
  \nonumber
\end{eqnarray}
Due to our two-zero-texture mass matrix (Eq.~(\ref{eq:nmzmNMM}))
we have an inverted hierarchy neutrino mass spectrum
\cite{Frampton:2002yf} and therefore 
$M_{ee}\simeq M_{\mu\tau}$. Thus the neutrino mass matrix, 
Eq.~(\ref{eq:nmzmNMM}), has only three independent
entries which, from Eqs.~(\ref{eq:neutrinoexpquant}),
can be written in terms of $\tan^{2}\theta_{23}$, 
$\tan 2\theta_{12}$ and $\Delta m_{23}^{2}$,
namely
\begin{eqnarray}
  \label{eq:neutrinoMMentries}
  M_{ee} &\simeq& M_{\mu\tau}\simeq\frac{\sqrt{\Delta m^2_{23}}}
  {\kappa\tan2\theta_{12}}\,,\nonumber\\
  M_{e\mu} &\simeq& \frac{\sqrt{2\Delta m^2_{23}}}
  {\kappa(1+\tan\theta_{23})}\,,\\
  M_{e\tau} &\simeq& -\frac{\tan\theta_{23}\sqrt{2\Delta 
      m^2_{23}}}{\kappa(1+\tan\theta_{23})}\,.\nonumber
\end{eqnarray}
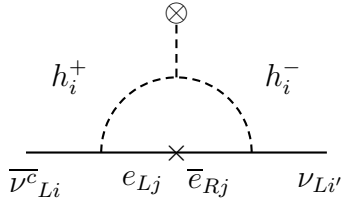
\begin{figure}[t]
  \begin{center}
    \begin{pspicture}(-5,1)(5,3)
      \psline(-2,1)(0,1)
      \uput[u](0,0.68){$\times$}
      \psline(0,1)(2,1)
      \psline[linestyle=dashed, dash=3pt 2pt](0,2.7)(0,2)
      \uput[u](0,2.54){$\otimes$}
      \psarc[linestyle=dashed, dash=3pt 2pt](0,1){1cm}{0}{180}
      \uput[r](1,2){$h_{i}^{-}$}
      \uput[l](-1,2){$h_{i}^{+}$}
      \uput[r](-2.4,0.6){$\overline{\nu^{c}}_{Li}$}
      \uput[l](2.4,0.6){$\nu_{Li'}$}
      \uput[r](-0.9,0.6){$e_{Lj}$}
      \uput[l](0.9,0.6){$\overline{e}_{Rj}$}
    \end{pspicture}
  \end{center}
  \caption{Loop diagrams for Majorana neutrino mass. Here $i=1,2$}
  \label{fig:loops}
\end{figure}
Assuming that there are no large hierarchies 
among the couplings $O_{i1}$ and
$O_{ij}$, terms proportional to $m_e$, in the 
neutrino mass matrix, (see
Eq.~(\ref{eq:massmatrix})) can be neglected. Thus we obtain
Eq.~(\ref{eq:nmzmNMM}) with $O_{23}=O_{32}=0$. Under this constraint
the mass matrix depends on $\kappa$ and on the seven parameters
\begin{equation}
  \label{eq:relevantparameters}
f_{12},f_{13},f_{23},O_{12},
O_{13},O_{22},O_{33},
\end{equation}
as can be seen from Eqs.~(\ref{eq:massmatrix}) and 
(\ref{eq:nmzmNMM}). By using equations in 
(\ref{eq:neutrinoMMentries}) we can write 
four of these parameters in terms of the other three. 
Equations (\ref{eq:a1}), (\ref{eq:a2}), (\ref{eq:a3})
and (\ref{eq:a4}), in the appendix, give the expressions for 
$f_{12}$, $f_{13}$, $f_{23}$, $O_{13}$, 
in terms of $O_{12}$, $O_{22}$, 
and $O_{33}$. Note that both $f_{23}$ and $O_{33}$ must be different 
from zero. 

Next we will consider the cases for which Eqs. (\ref{eq:a1}),
(\ref{eq:a2}), (\ref{eq:a3}) and (\ref{eq:a4}) can be expressed in
terms of a single parameter. We will call these cases the
one-parameter solutions. Since $O_{33}$ cannot be zero 
(see Eq.~(\ref{eq:a2})) we will
parametrize all our one-parameter solutions in terms of this
coupling. This leaves us with only four possibilities: 
$O_{12}=0$, $O_{13}=0$, $f_{12}=0$, $f_{13}=0$
and the remaining parameters in Eq.~(\ref{eq:relevantparameters})
different from zero in each case. We will show 
below that the first two lead to solutions with 
large $O_{33}$, while the last two lead to 
solutions with small $O_{33}$.
\subsection{The one-parameter region}
\label{sec:oneparameterregion}
The main point here is that in these two cases (small and 
large $O_{33}$) not only neutrino physics but the 
decay patterns of $h^{\pm}_{1}$ are governed by 
a single parameter. This allows an analytical approach 
to the problem of identifying a particular collider 
signature that allows to distinguish between different regions in
parameter space. In the following we will discuss the four
possibilities mentioned previously and we will estimate
the values of the parameters, consistent with neutrino physics
as well as with FCNC constraints, in each case. This discussion
will be useful in our analysis of the decays of $h_{1}^{+}$ 
presented in section \ref{sec:hierarchy}.
\subsubsection{The large $O_{33}$ case}
\label{sec:large-o_33}
Choosing $O_{12}=0$ and $O_{22}=(m_\mu/m_\tau)O_{33}$,
as in references \cite{Balaji:2001ex,Hasegawa:2003by}, 
Eqs. (\ref{eq:a1}), (\ref{eq:a2}), (\ref{eq:a3}) and 
(\ref{eq:a4}) are reduced to the one-parameter solution
\begin{align}
  f_{12}\approx&\frac{\left[ 1 + 
      \left( 2 + 4\,\tan^2 2\theta_{12}
      \right) \,\tan\theta_{23} + 
      \tan^2\theta_{23} \right]\,}{2\,
    \sqrt{2}\,\kappa \,\tan^2 2\theta_{12}\,
    \tan\theta_{23}\,
    \left( 1 + \tan\theta_{23} \right)}
  \frac{\sqrt{\Delta m_{23}^2}\,m_\tau}{m_\mu^2}
  \frac{1}{O_{33}}\sim\frac{6.3\times10^{-9}}{\kappa\,O_{33}}, \nonumber\\
  \label{eq:1}
  f_{13}\approx&-\frac{\sqrt{2}\,\tan\theta_{23}}{\kappa \,
    \left( 1 + \tan\theta_{23} \right) }
  \frac{\sqrt{\Delta m_{23}^2}}{m_{\tau }}
  \frac{1}{O_{33}\,}\sim-\frac{1.9\times10^{-11}}{\kappa\,O_{33}},
  \\
  f_{23}\approx&\frac{1}
  {\kappa \,\tan2\theta_{12}}
  \frac{\sqrt{\Delta m_{23}^2}}{m_{\tau }}
  \frac{1}{O_{33}\,}\sim\frac{1.2\times10^{-11}}{\kappa\,O_{33}},
  \nonumber\\
  O_{13}\approx&-\frac{  1 + \tan\theta_{23}    }{2\,
    \sqrt{2}\,\tan 2\theta_{12}\,\tan\theta_{23}}O_{33}\sim-0.3\,O_{33}.
  \nonumber
\end{align}
The last values in each equation are obtained using the best fit 
point value for each neutrino observable. 

An upper bound for $\kappa$ can be estimated using the fact that
\begin{equation}
  \label{eq:2}
   \kappa=  \frac{\sin2\varphi}{(4\pi)^{2}}
  \ln\left(\frac{M^{2}_{2}}{M^{2}_{1}}\right)=
\frac{\sqrt{2}v\mu}{(4\pi)^{2}}
  \frac{1}{M_{2}^2 -
    M_{1}^2}\ln\left(\frac{M^{2}_{2}}{M^{2}_{1}}\right)\simeq
  \frac{\sqrt{2}}{(4\pi)^{2}}\frac{v\mu}{M_2^2}.
\end{equation}
Therefore for $M_2<1000\,$GeV and $|\mu|<500\,$GeV \cite{Barroso:2005hc}, 
we have that $|\kappa|\lesssim10^{-2}$.  For example for $M_1=200\,$GeV,
$M_2=300\,$GeV, and $\mu=100\,$GeV, we have
\begin{equation}
  \label{eq:3}
  \sin2\varphi=0.7\qquad\text{and}\qquad \kappa=3.6\times10^{-3}.
\end{equation}
On the other hand, from the expression for $f_{12}$ in
Eq.~\eqref{eq:1} and imposing $f_{12}\lesssim10^{-2}$
a lower bound on $\kappa$ can be found. Choosing 
$O_{33}\lesssim10^{-2}$, we have that $|\kappa|\gtrsim10^{-5}$.
For example, for $\mu=2\,$GeV and with $M_1$ and $M_2$ as in the 
previous case, we have
\begin{equation}
  \label{eq:4}
  \sin2\varphi=0.014\qquad\text{and}\qquad \kappa=7.2\times10^{-5}.
\end{equation}
Using the value of $\kappa$ given in Eq.~\eqref{eq:3} we have
\begin{equation}
  \label{eq:caseb}
  f_{12}\sim\frac{1.8\times10^{-6}}{O_{33}}, \qquad
  f_{13}\sim-\frac{5.2\times10^{-9}}{O_{33}},\qquad
  f_{23}\sim\frac{3.2\times10^{-9}}{O_{33}}.\qquad
\end{equation}
Now instead of $O_{12}=0$ we choose $O_{13}=0$. Again, as in 
the previous case, we take $O_{22}=(m_\mu/m_\tau)O_{33}$ and the 
best fit point values for each neutrino observable. With $\kappa$ 
given by \eqref{eq:3}, Eqs.~(\ref{eq:a1}), (\ref{eq:a2}), 
(\ref{eq:a3}) and (\ref{eq:a4}) become
\begin{equation}
  \label{eq:casec}
  f_{12}\sim\frac{1.5\times10^{-6}}{O_{33}},\;
  f_{13}\sim-\frac{5.2\times10^{-9}}{O_{33}},\;
  f_{23}\sim\frac{3.2\times10^{-9}}{O_{33}},\;
  O_{13}\sim0.02\,O_{33},
\end{equation}
which is basically the same result obtained in the case 
with $O_{12}=0$ (Eqs.~(\ref{eq:caseb})).

From Eqs.~(\ref{eq:caseb}) and (\ref{eq:casec}) it can be
seen that all the parameters can be below $10^{-3}$,
with a hierarchy of order $10^{3}$ between
$f_{12}$ and the others $f_{ij}$. In this way the constraints
on the couplings coming from FCNC interactions 
(Tables~\ref{tab:fcnch0} and \ref{tab:fcnchc}) are always
satisfied. 

Note that $f_{12}\lesssim10^{-2}$ requires
$O_{33}\gtrsim10^{-4}$. Therefore the range of variation of 
$O_{33}$ is restricted to $10^{-4}\lesssim O_{33}\lesssim10^{-2}$. 
For $\kappa$ small, as in Eq.~\eqref{eq:4},  $O_{33}\sim10^{-2}$.
\subsubsection{The small $O_{33}$ case}
\label{sec:small-o_33}
If we choose $f_{13}=0$ and, in order to define the one-parameter 
solution in this case\footnote{This choice allow us to
define the one-parameter solutions. However, we stress that
our results does not depend on this choice.
Our main conclusions hold for any $O_{22}<O_{33}$.}
$O_{22}=0$, Eqs.~(\ref{eq:a1}), (\ref{eq:a2}), (\ref{eq:a3}) 
and (\ref{eq:a4}) become
\begin{align}
  f_{12}\approx&\frac{\left( 1 + \tan\theta_{23} \right)}{2\,\sqrt{2}\,
    \kappa \,
    \tan2\theta_{12}^2\,
    \tan\theta_{23}}\frac{\sqrt{\Delta m_{23}^2}}{m_\tau \,}
  \frac{1}{O_{33}}\sim\frac{3.6\times10^{-12}}{\kappa\,O_{33}},\nonumber\\
  \label{eq:5}
  f_{23}\approx&\frac{1}
  {\kappa \,
    \tan2\theta_{12}}\frac{\sqrt{\Delta m_{23}^2}}{m_\tau}\frac{1}
  {O_{33}}\sim\frac{1.2\times10^{-11}}{\kappa\,O_{33}},\\
  O_{12}\approx&\frac{{\sqrt{2}}\,
    \tan2\theta_{12}\,
    \tan\theta_{23}}{
    \left( 1 + \tan\theta_{23} \right) }\frac{m_\tau}{m_\mu}O_{33}
  \sim 27\,O_{33},
  \nonumber\\
  O_{13}\approx&\frac{\sqrt{2}\,\tan2\theta_{12}}
  {1 + \tan\theta_{23}}O_{33}\sim1.6\,O_{33}.
  \nonumber
\end{align}
The last values in each equation are obtained using the best fit 
point values for each neutrino observable. Note that
$O_{12}\lesssim10^{-2}$ requires $O_{33}\lesssim4\times10^{-4}$. 

On the other hand, from the expression for $f_{23}$ in
Eq.~\eqref{eq:5}, if $O_{33}\lesssim4\times10^{-4}$ and 
we impose the bound $f_{23}\lesssim 10^{-2}$ we have that 
$\kappa\gtrsim3\times10^{-6}$. For example, if we choose  
$\mu=0.2\,$GeV, $M_1=200$\,GeV and $M_2=300$\,GeV , we have
\begin{equation}
  \label{eq:6}
  \sin2\varphi=1.4\times10^{-3}\qquad\text{and}\qquad 
  \kappa=7.2\times10^{-6}
\end{equation}

For the value of $\kappa$ given in Eq.~\eqref{eq:3}, that 
satisfies the bound $\kappa\gtrsim3\times10^{-6}$,
with $M_1=200\,$ GeV and $M_2=300\,$GeV we have
\begin{equation}
  f_{12}\sim\frac{1\times10^{-9}}{O_{33}},\quad
  f_{23}\sim\frac{3.2\times10^{-9}}{O_{33}}.
\end{equation}
Now instead of $f_{13}=0$ we choose $f_{12}=0$.
Using the best fit point values for each
neutrino observable, $O_{22}=0$ and $\kappa$ given by Eq.~\eqref{eq:3}, 
Eqs.~(\ref{eq:a1}), (\ref{eq:a2}), (\ref{eq:a3}) and (\ref{eq:a4})
become
\begin{equation}
  f_{13}\sim\frac{9.8\times10^{-10}}{O_{33}},\;
  f_{23}\sim\frac{3.2\times10^{-9}}{O_{33}},\;
  O_{12}\sim 32\,O_{33},\;
  O_{13}\sim1.6\,O_{33}.
\end{equation}
In both cases ($f_{12}=0$ or $f_{13}=0$) we can have all the 
five parameters of order of $10^{-4}$ without any hierarchy among 
them. In fact, the case $f_{12}=0$, considered here, is a particular 
case of the one studied in reference \cite{He:2003ih} in which  
all the parameters $O_{ij}$ and $f_{ij}$ are of the same order of
magnitude. From Tables~\ref{tab:fcnch0} and \ref{tab:fcnchc}
it can be seen that FCNC constraints are always satisfied. 
\begin{table}[t]
  \centering
  \begin{tabular}{|l||c||c||c|}\hline\hline
    Case         &$O_{33}$&$\kappa$&$\mu$ (GeV)\\\hline
    Large $O_{33}$&$10^{-4}$ -- $10^{-2}$&$10^{-5}$ -- $10^{-2}$
    &2 -- 500\\
    Small $O_{33}$&$10^{-7}$ -- $4\times10^{-3}$
    &$3\times10^{-6}$ -- $10^{-2}$&$0.2$ -- 500\\\hline\hline
  \end{tabular}
  \caption{Range of $O_{33}$, $\kappa$ and 
    $\mu$ for the one-parameter solutions in the NMZM}
  \label{tab:1}
\end{table}

A lower bound on $O_{33}$ can be obtained using the bound
$f_{23}\lesssim10^{-2}$. Together with the upper bound estimated
previously we have $10^{-7}\lesssim
O_{33}\lesssim4\times10^{-3}$. Notice that for smaller values of
$\kappa$, as the one in Eq.~\eqref{eq:6}, the range of variation is 
more restricted, $10^{-5}\lesssim O_{33}\lesssim4\times10^{-3}$.

It is worth noticing that there are no more possibilities in 
the one-parameter solution case. The large $O_{33}$ case, 
obtained when either $O_{12}$ or $O_{13}$ are neglected 
implies a hierarchy among the non zero $f_{ij}$ and, 
depending on the case, on $O_{12}$ or $O_{13}$. In the 
small $O_{33}$ case, obtained when either 
$f_{12}$ or $f_{13}$ are neglected, it is possible
to have all the parameters at the level of $10^{-4}$.
Table~\ref{tab:1} shows the allowed range of variation for
$O_{33}$, $\kappa$ and $\mu$ in each case.
\section{Neutrino and Collider Physics}
\label{sec:NMZMandCP}
\subsection{Determination of the Neutrino Mass Matrix Parameters}
\label{sec:detNMMparam}
In this section we discuss how the charged scalar decays
can give some hints about the parameters that determine
neutrino masses and mixing angles. Charged Higgs decays are 
governed by the same parameters that control neutrino physics so,
in principle, the information coming from these decays
can be used to reconstruct the neutrino mass matrix.
Outside of the one-parameter regions analysed in section
\ref{sec:neutrinophysics} the number of parameters is large and 
since neutrino flavour cannot be determined the mass matrix
cannot be, in general, reconstructed. Despite
this, in the limiting case of small mixing ($\varphi\ll 1$),
the fact that the the mainly doublet state decays are dictated by
the $O_{ij}$ and the mainly singlet state decays
are controlled by the $f_{ij}$ leads to a
situation in which the reconstruction of
part of the parameter space of the model is possible.

The charged scalar singlet $h^{+}$ does not couples to quarks. 
Thus experimentally the mainly singlet state can be 
differentiated from the mainly doublet state by the fact 
that the branching ratio to final states with quarks 
($\bar{u}_{i}d_{j}$) must be smaller for the former than 
for the latter. Our main assumption here is that all the decays 
that we are going to consider have a branching ratio 
in the order of at least per-mille. 

In the following discussion we will use the notation $h^{+}_{d,s}$ for 
charged Higgses. Here $d$ and $s$ denote the mainly doublet
and mainly singlet states respectively. Note that $d=1\;,s=2$
or $d=2\;,s=1$ are possible. Ratios of branching ratios 
for $h_{d,s}^{+}$ can be used
to obtain information about the $O_{ij}$ and $f_{ij}$
couplings. In the case of $h_{d}^{+}$ we have 
\begin{align}
  \label{eq:rofbrratios1}
  \frac{Br(h_{d}^{+}\to(\sum_{i}\nu_{i})\ell^{+}_{j})}
  {Br(h_{d}^{+}\to(\sum_{k}\nu_{k})\ell^{+}_{s})}\simeq
  \frac{\sum_{i}O_{ij}^{2}}{\sum_{k}O_{ks}^{2}}
\end{align}
and for $h_{s}^{+}$
\begin{align}
  \label{eq:rofbrratios2}
  \frac{Br(h_{s}^{+}\to(\sum_{i}\nu_{i})\ell^{+}_{j})}
  {Br(h_{s}^{+}\to(\sum_{k}\nu_{k})\ell^{+}_{s})}\simeq
  \frac{\sum_{i}f_{ij}^{2}}{\sum_{k}f_{ks}^{2}}.
\end{align}
Corrections to both ratios are $\propto\varphi^2\lll 1$.
The interesting point here is that despite the large number
of parameters the relative size of the $f_{ij}$ couplings
can be obtained by suitable combinations of ratios of 
branching ratios, for example
\begin{equation}
  \label{eq:fijsize}
  \frac{Br^{\mu}_s-Br^{\tau}_s+Br^{e}_s}
  {Br^{\mu}_s-Br^{e}_s+Br^{\tau}_s}\simeq
  \frac{f_{12}^{2}}{f_{23}^{2}}
\end{equation}
with $Br^{\ell_{j}}_s$ denoting
$Br(h_{s}^{+}\to(\sum_{i}\nu_{i})\ell^{+}_{j})$.
For the $O_{ij}$ the situation is more complicated but
even in this case some information can be obtained 
from the ratios of branching ratios. For example, the relation
\begin{equation}
  \label{eq:oijsize}
  \frac{Br(h_{d}^{+}\to(\sum_{i}\nu_{i})\mu^{+})}
  {Br(h_{d}^{+}\to(\sum_{k}\nu_{k})e^{+})}\simeq
  \frac{O_{12}^{2}+O_{22}^{2}}
  {O_{11}^{2}+O_{21}^{2}+O_{31}^{2}}
\end{equation}
allows to determine the relative importance of the 
couplings involved in these decays.

\begin{figure}[t]
  \centering
  \begin{pspicture}(-7,-1)(7,5)
    \uput[r](-6.2,2)
    {\mbox{\includegraphics
        [width=5.8cm, height=5cm]{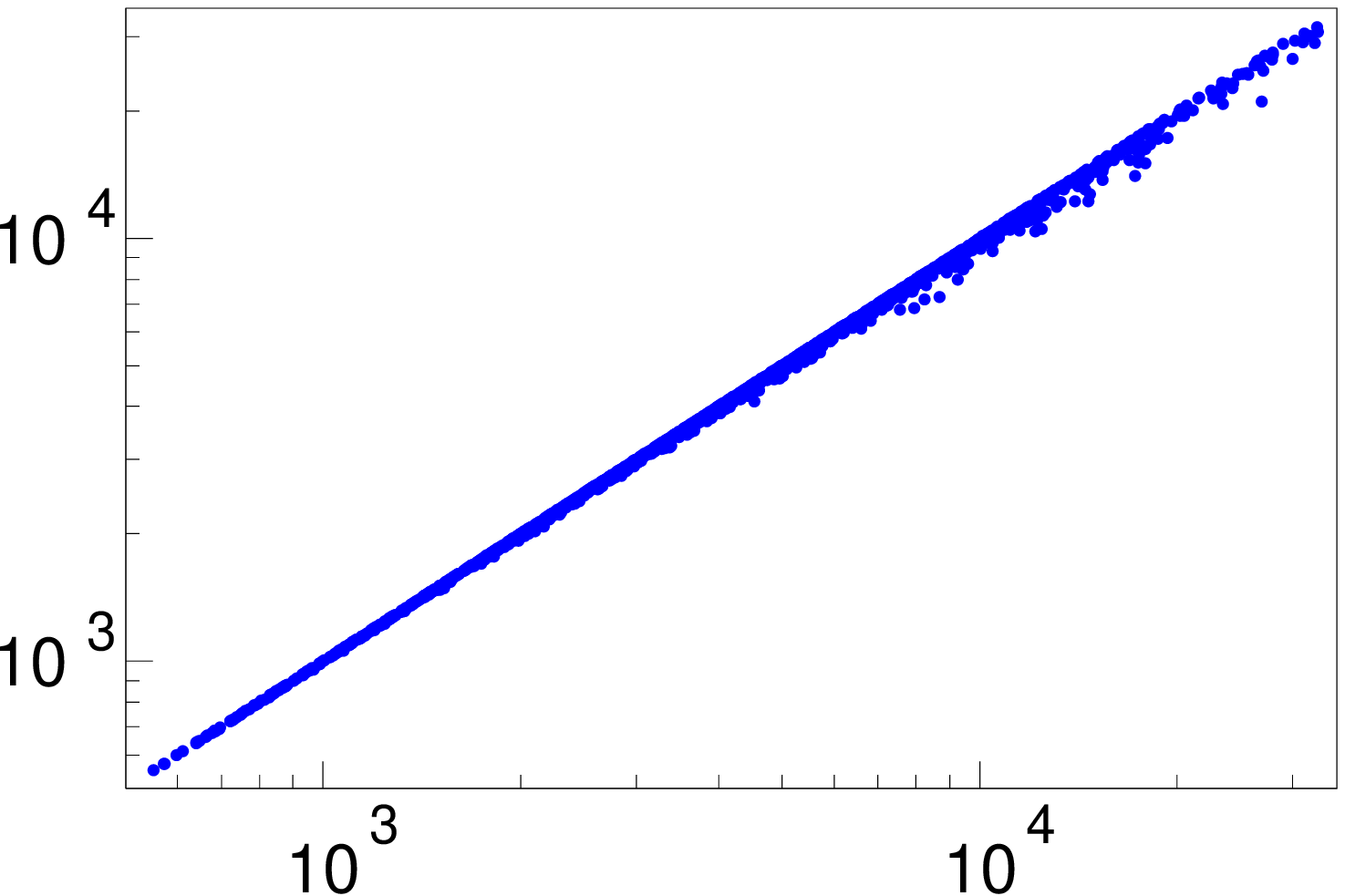}}}
    \uput[r](0.4,2)
    {\mbox{\includegraphics
        [width=5.8cm, height=5cm]{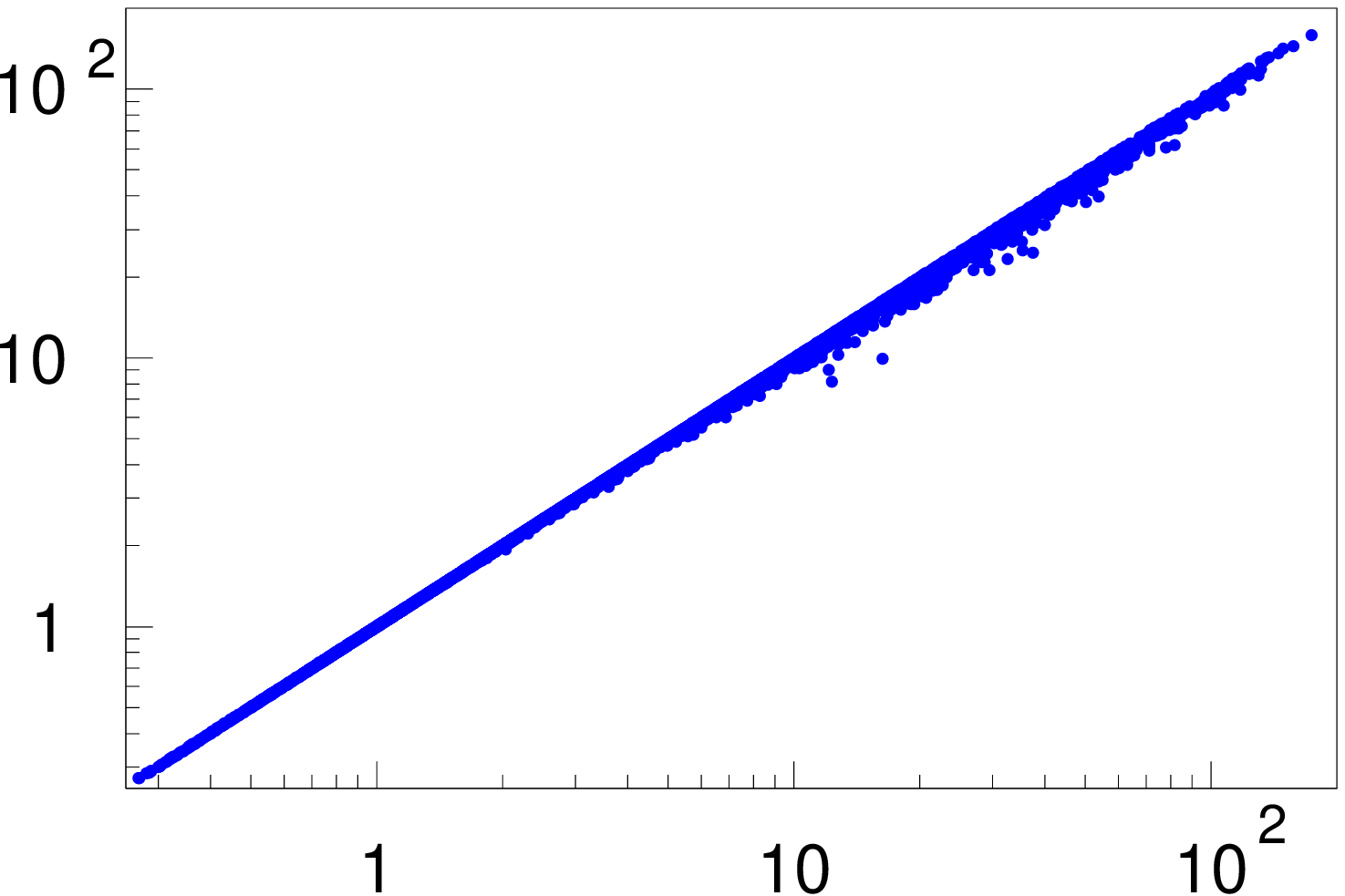}}}
    \rput{W}(-6.5,3.2){$Br^{\mu\tau e}_s/Br^{\mu e\tau}_s$}
    \uput[r](-1.7,-0.9){$f_{12}^{2}/f_{23}^{2}$}
    \rput{W}(0.2,3.5){$Br^{\mu}_d/Br^{e}_d$}
    \uput[r](4.9,-0.9){$O_{\mu}^{2}/O_{e}^{2}$}
  \end{pspicture}
  \caption{Ratio of branching ratios 
    $Br^{\mu\tau e}_s/Br^{\mu e\tau}_s=
    (Br^{\mu}_s-Br^{\tau}_s+Br^{e}_s)/
    (Br^{\mu}_s-Br^{e}_s+Br^{\tau}_s)$ versus 
    $f_{12}^{2}/f_{23}^{2}$ (left) 
    and $Br^{\mu}_d/Br^{e}_d=
    Br(h_{d}^{+}\to(\sum_{i}\nu_{i})\mu^{+})/
    Br(h_{d}^{+}\to(\sum_{k}\nu_{k})e^{+})$ versus
    $O_{\mu}^{2}/O_{e}^{2}=
    (O_{12}^{2}+O_{22}^{2})/(O_{11}^{2}+O_{21}^{2}+O_{31}^{2})$
    (right). See text.}
  \label{fig:correlation}
\end{figure}
Figure~\ref{fig:correlation} shows the ratios of 
branching ratios described above. Any deviation 
from the small mixing assumption would lead to a 
large dispersion. 

\begin{figure}[t]
  \centering
  \begin{pspicture}(-7,-1)(7,5)
    \uput[r](-6.2,2)
    {\mbox{\includegraphics
        [width=5.8cm, height=5cm]{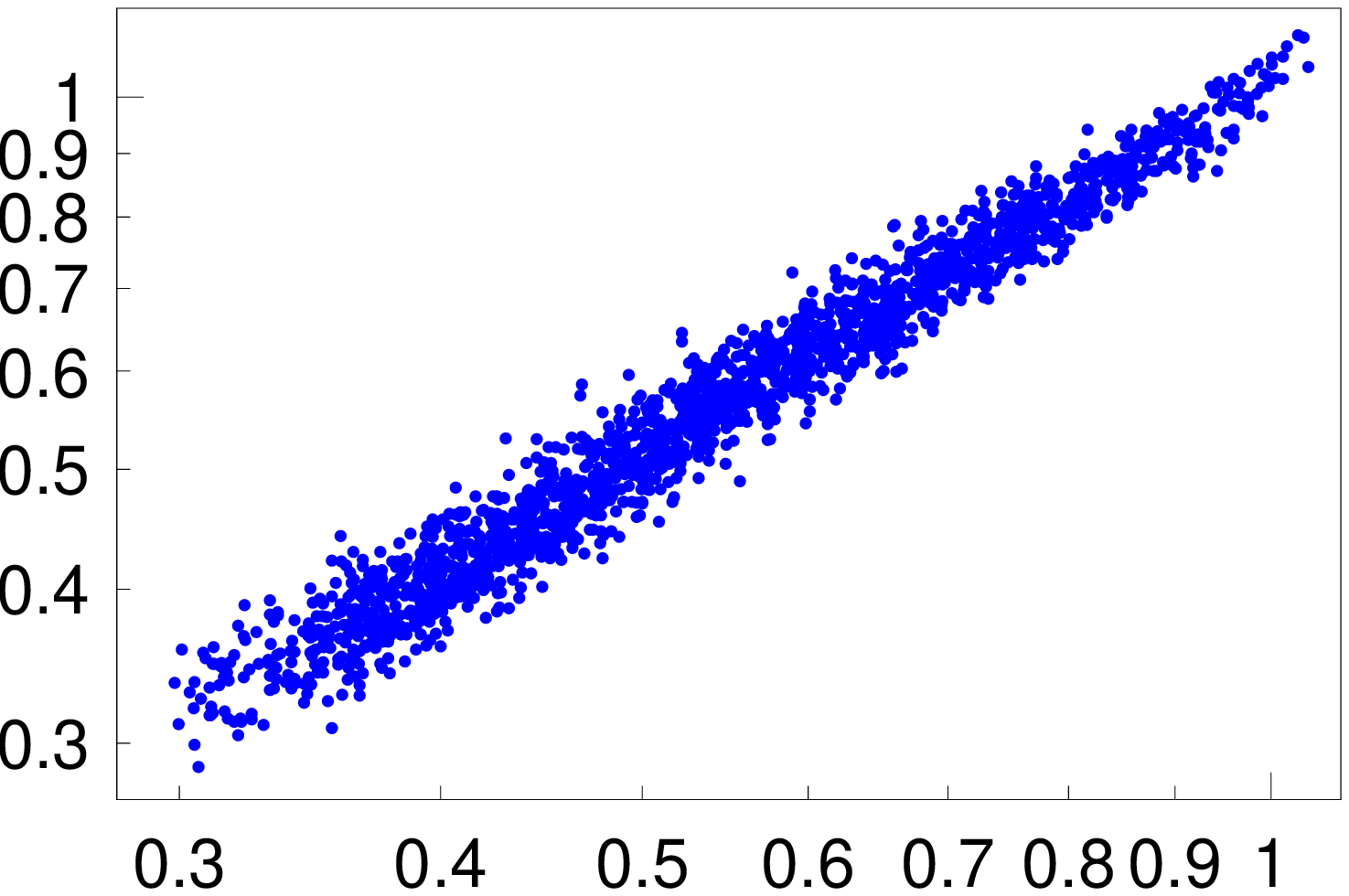}}}
    \uput[r](0.4,2)
    {\mbox{\includegraphics
        [width=5.8cm, height=5cm]{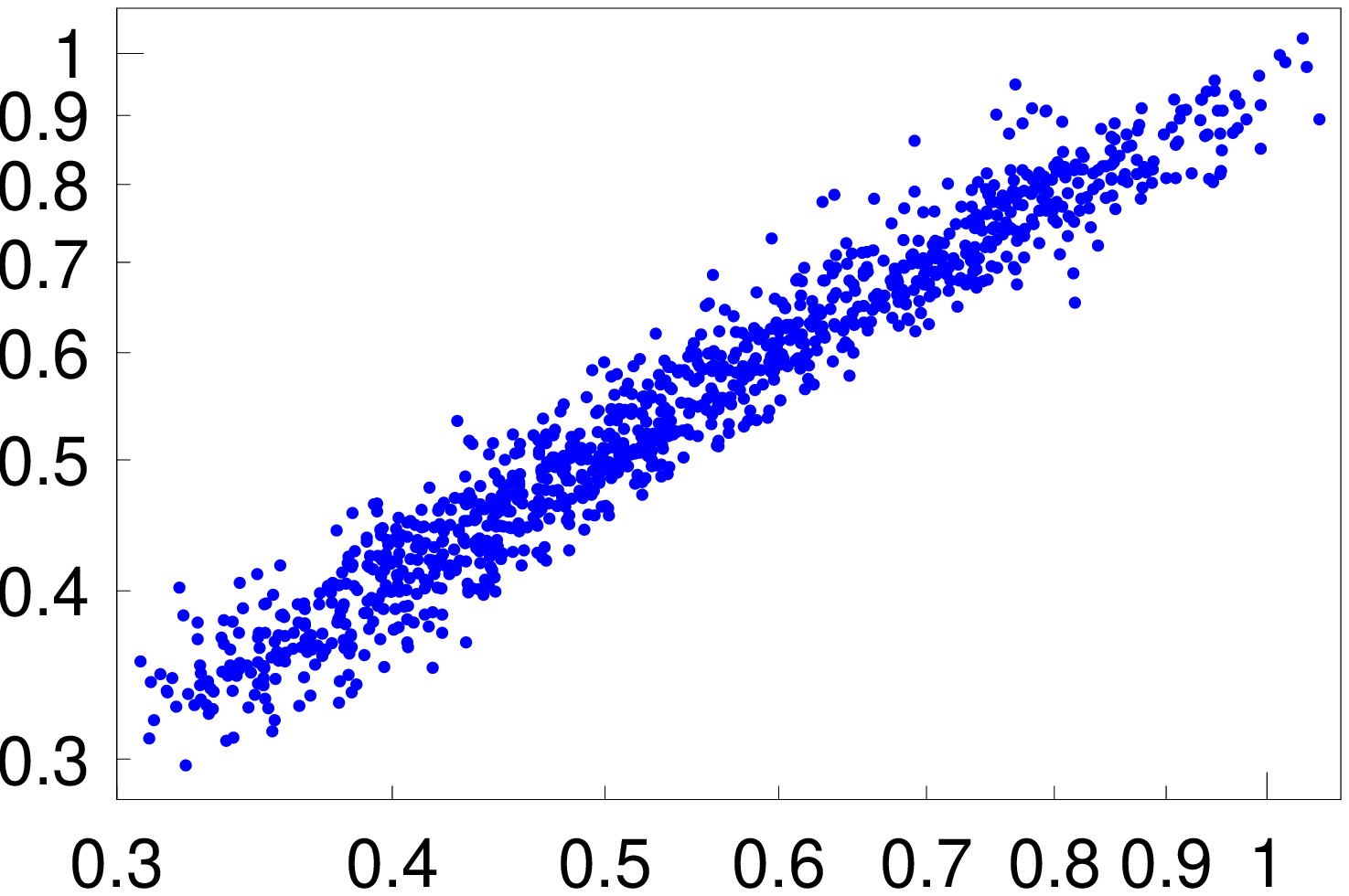}}}
    \rput(-6.3,4.5){$y_1$}
    \uput[r](-0.8,-0.8){$x_1$}
    \rput(0.4,4.5){$y_2$}
    \uput[r](5.8,-0.9){$x_2$}
  \end{pspicture}
  \caption{Ratio of branching ratios indicated by the variables
    $y_1$ (left) and $y_2$ (right) versus the atmospheric and solar
    mixing angles indicated by the variables $x_1$ (left) and $x_2$
    (right). See text}
  \label{fig:correlationmixa}
\end{figure}

There are two limit cases of particular interest where the 
decays of $h_{d,s}^{+}$ are correlated with the neutrino mixing
angles, $O_{12}\ll O_{13} \ll O_{22}<O_{33}$ or 
$O_{13}\ll O_{12}\ll O_{22}<O_{33}$. 
Figure~\ref{fig:correlationmixa} shows both cases. In the left
plot the variables $y_1$ and $x_1$ are given by
\begin{align}
  \label{eq:variable1}
   y_{1}&=
   \sqrt{\frac{Br^{\mu}_s-Br^{e}_s+Br^{\tau}_s}
   {Br^{e}_s-Br^{\mu}_s+Br^{\tau}_s}}
   \left(
     1-\frac{m_{\mu}}{m_{\tau}}
     \sqrt{
       \frac{Br(h_{d}^{+}\to(\sum_{i}\nu_{i})\mu^{+})}
       {Br(h_{d}^{+}\to(\sum_{k}\nu_{k})\tau^{+})}}
   \right)\nonumber\\
   x_1&=\frac{1}{\sqrt{2}\tan2\theta_{12}}
   \left(
     1+\frac{1}{\tan\theta_{23}}
   \right).
\end{align}
In the right one the variables $y_2$ and $x_2$ are defined
as
\begin{align}
  \label{eq:variables2}
  y_2&=
   \sqrt{\frac{Br^{\mu}_s-Br^{e}_s+Br^{\tau}_s}
   {Br^{\mu}_s-Br^{\tau}_s+Br^{e}_s}}
   \left(
     \frac{m_{\tau}}{m_{\mu}}
     \sqrt{
       \frac{Br(h_{d}^{+}\to(\sum_{i}\nu_{i})\tau^{+})}
       {Br(h_{d}^{+}\to(\sum_{k}\nu_{k})\mu^{+})}}
     -1
   \right)\nonumber\\
   x_2&=\frac{1}{\sqrt{2}\tan2\theta_{12}}
   \left(
     1+\tan\theta_{23}
   \right)
\end{align}
\begin{figure}[t]
  \centering
    \begin{pspicture}(-6,-1)(4,6)
    \uput[r](-5.2,2.3)
    {\mbox{\includegraphics
        [width=9cm, height=6cm]{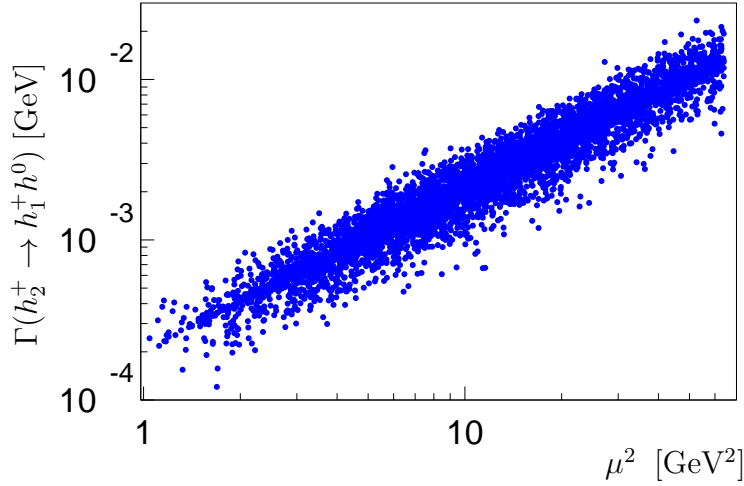}}}
    \rput{W}(-5.5,2.5){
      $\Gamma(h_{2}^{+}\rightarrow h^{+}_{1}h^{0})\;$[GeV]}
    \uput[r](2,-0.9){$\mu^{2}\;\;$[GeV$^2$]}
  \end{pspicture}
  \caption{Decay rate 
    $\Gamma(h_{2}^{+}\rightarrow h^{+}_{1}h^{0})$ versus
  $\mu^{2}$ for fixed values 
  $M_{2}=400\,\text{Gev}$, $M_{1}=150\,\text{Gev}$
  and $M_{h^{0}}=130\,\text{GeV}$.}
  \label{fig:h2h1h0}
\end{figure}
Another important decay, if kinematically allowed, that could
be used to obtain information about $\mu$ is
$h_{2}^{+}\rightarrow h^{+}_{1}h^{0}$. The decay rate for this
process reads
\begin{equation}
  \label{eq:decaywh2h0h1}
  \Gamma(h_{2}^{+}\rightarrow h^{+}_{1}h^{0})=
  \frac{1}{16\pi}\frac{\Lambda^{2}}{M_{2}}
  \sqrt{1-4\frac{M_{1}^{2}}{M_{2}^{2}}}.
\end{equation}
Here
\begin{equation}
  \label{eq:lamdacoupling}
  \Lambda=\frac{\mu}{\sqrt{2}}\sin\alpha\cos2\varphi
  +v\frac{\sin2\varphi}{2}(\Lambda_{22}-\Lambda_{33})
\end{equation}
and
\begin{align}
  \label{eq:lamda22andlamda33}
  \Lambda_{22}&=\lambda_{7}\cos\alpha-\lambda_{3}\sin\alpha,\nonumber\\
  \Lambda_{33}&=\lambda_{10}\cos\alpha-\lambda_{8}\sin\alpha
\end{align}
where $\alpha$ is the mixing angle that define the 
two CP-even Higgs mass eigenstates, $h^{0}$ and $H^{0}$.

Figure~\ref{fig:h2h1h0} shows the decay rate 
$\Gamma(h_{2}^{+}\rightarrow h^{+}_{1}h^{0})$
versus $\mu^{2}$. There we have fixed $M_{2}=400\,\text{Gev}$, 
$M_{1}=150\,\text{Gev}$, $M_{h^{0}}=130\,\text{GeV}$ and 
$\alpha=\pi/6$. $\mu$ is in the range $0.1\,\text{GeV}-8\,\text{Gev}$ 
in order to ensure $\varphi\ll 1$. The dispersion is due to the presence
of the other couplings, present in the scalar potential 
(Eq.~(\ref{eq:scalarpotentialinhiggsbas})). Apart from allowing
the approximate determination of $\mu$, measurements
of $\Gamma(h_{2}^{+}\rightarrow h^{+}_{1}h^{0})$ in
the range indicated by Fig.~\ref{fig:h2h1h0} will indicate
that the small mixing limit is realized.
\subsection{Hierarchy of charged Higgs leptonic decays}
\label{sec:hierarchy}
\begin{figure}
  \centering
  \includegraphics[width=8cm, height=7.5cm]{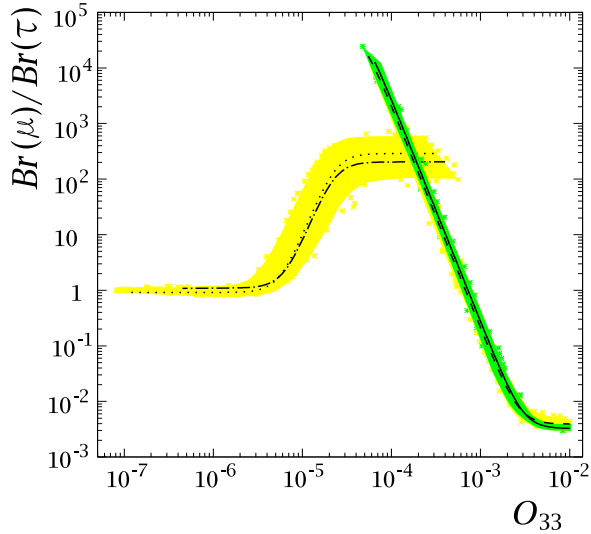}  
  \caption{${Br(h_{1}^{+}\to(\sum_{i}\nu_{i})\mu^{+})}/
    {Br(h_{1}^{+}\to(\sum_{i}\nu_{i})\tau^{+})}$ as 
    a function of $O_{33}$ for $3\times10^{-6}<\kappa<10^{-2}$ 
    obtained with $0.2<\mu<500\,$GeV, $M_1=200\,$GeV
    and $M_{2}=500\,$GeV. All the parameters $f_{ij}$ and 
    $O_{ij}$ satisfy the bounds shown in Tables~\ref{tab:fcnch0}
    and \ref{tab:fcnchc}. For all the dark gray
    (green) points $O_{12}<10^{-6}$. See text}
  \label{fig:BR-O33}
\end{figure}
In this section we will show that in the NMZM, the
decay process $h_{1}^{+}\to (\sum_{i}\nu_{i})\mu^{+}$
is enhanced in comparision to the 2HDM of type-I and 
type-II. Moreover, it is shown that in large parts of
the parameter space, $h_{1}^{+}\to (\sum_{i}\nu_{i})\mu^{+}$
can be the dominant leptonic decay.

In the one-parameter solutions, described in
sec.~\ref{sec:neutrinophysics}, the parameters
$O_{12}$, $O_{13}$, $f_{12}$, $f_{13}$ and $f_{23}$
are governed by the parameter $O_{33}$. In this way the 
$Br(h^{+}_{1}\rightarrow(\sum_{i}\nu_{i})\ell_{j}^{+})$
are functions of $O_{33}$.
In order to find expressions with no dependence on 
$\kappa$ or $O_{33}$ and correlated with 
neutrino physics observables 
($\tan 2\theta_{12}$, $\tan \theta_{23}$)
we consider ratios of branching ratios in the limits
$O_{ij}\gg f_{ij}$ and  $O_{ij}\ll f_{ij}$, namely
\begin{align}
  \label{eq:brratio}
  \frac{Br(h_{1}^{+}\to(\sum_{i}\nu_{i})\mu^{+})}
  {Br(h_{1}^{+}\to(\sum_{i}\nu_{i})\tau^{+})}=&
  \frac{\sum_{i}[(O_{i2}\cos\varphi)^{2} + 
    (2f_{i2}\sin\varphi)^{2}]}{\sum_{i}[(O_{i3}\cos\varphi)^{2} + 
    (2f_{i3}\sin\varphi)^{2}]}\\
  \label{eq:brratioapprox}
  \approx&
  \begin{cases}
    \displaystyle{\frac{\sum_{i}O_{i2}^{2} 
    }{\sum_{i}O_{i3}^{2}}}& \text{for $O_{ij}\gg f_{ij}$}\\
    \displaystyle{\frac{\sum_{i}f_{i2}^{2} 
    }{\sum_{i}f_{i3}^{2}}}& \text{for $O_{ij}\ll f_{ij}$}
  \end{cases}
\end{align}
Clearly from Eqs.~\eqref{eq:1} or \eqref{eq:5},
Eq.~\eqref{eq:brratioapprox} depend only on the neutrino mixing angles
and charged lepton masses. We will call the regions 
of parameter space with either
$O_{ij}$ or $f_{ij}$ dominance correlation
regions. Ratio of branching ratios in these regions are $\kappa$ 
independent (or $\mu$ independent). In general, outside the correlation 
regions, the independence on $\mu$ approximately holds, 
but there is a dependence on $M_2$.

Fig.~\ref{fig:BR-O33} shows the ratio  
${Br(h_{1}^{+}\to(\sum_{i}\nu_{i})\mu^{+})}/
{Br(h_{1}^{+}\to(\sum_{i}\nu_{i})\tau^{+})}$ as function of $O_{33}$. 
For all curves, we have used the best fit point values for 
$\Delta m_{23}^{2}$, and the solar and atmospheric 
mixing angles. We have taken also $3\times10^{-6}<\kappa<10^{-2}$ 
obtained when $0.2<\mu<500\,$GeV, $M_1=200\,$GeV
and $M_{2}=500\,$GeV. The correlation regions correspond to the
flat parts of the curves. The
large $O_{33}$ case determined by Eq.~(\ref{eq:1}) 
with $O_{12}=0$ corresponds to the solid line in the
right part of the plot, while the large $O_{33}$ case with
$O_{13}=0$ correspond to the dashed line. In the same way, the small
$O_{33}$ case described by Eq.~(\ref{eq:5}) with $f_{13}=0$
corresponds to the dotted line in the left part of the plot, while the
small $O_{33}$ case with $f_{12}=0$ correspond to the
dotted--dashed line. The scatter plot was obtained by 
searching for all solutions compatible with neutrino 
data at 3$\sigma$ level, and keeping 
$O_{22}=(m_\mu/m_\tau)\,O_{33}$.

In the large $O_{33}$ case described by Eq.~(\ref{eq:1}),
the correlation region for $f_{ij}\gg O_{ij}$ is excluded 
because the parameters $f_{ij}$ are above the values 
consistent with FCNC constraints
(see Tables~\ref{tab:fcnch0} and \ref{tab:fcnchc}).  
For the other correlation region, for which 
$O_{ij}\gg f_{ij}$, we have
\begin{equation}
  \label{eq:largeO33}
  \frac{Br(h_{1}^{+}\to(\sum_{i}\nu_{i})\mu^{+})}
  {Br(h_{1}^{+}\to(\sum_{i}\nu_{i})\tau^{+})}\sim
  \left(
    \frac{m_\mu}{m_\tau}
  \right)^2\qquad\text{for $O_{ij}\gg f_{ij}$}.
\end{equation}
As shown by the solid line at the right of Fig.~\ref{fig:BR-O33}, the
contribution of $f_{ij}$ can increase the ratio of branchings ratios 
up to a factor of $10^{7}$. In this way the decay 
$h_{1}^{+}\rightarrow(\sum_{i}\nu_{i})\mu^{+}$
may become observable in future colliders. The 
dark gray (green) points were selected from the full scatter plot 
by choosing $O_{12}<10^{-6}$. They are well fitted by the solid line 
which represents the one-parameter solution with $O_{12}=0$ as given 
in Eq.~(\ref{eq:1}).

In the small $O_{33}$ case with $f_{13}=0$, we have
\begin{align}
&\frac{Br(h_{1}^{+}\to(\sum_{i}\nu_{i})\mu^{+})}
{Br(h_{1}^{+}\to(\sum_{i}\nu_{i})\tau^{+})}\nonumber\\
\label{eq:smallO33}
&\approx \begin{cases}
    \displaystyle{\frac{2\,
    \tan^22\theta_{12}\,
    \tan^2\theta_{23}}{
    2\,\tan^22\theta_{12} + 
      {\left( 1 + \tan\theta_{23} \right) }^2
       }\frac{m_\tau^2}{m_\mu^2\,}}&\text{for $O_{ij}\gg f_{ij}$}\\
    \displaystyle{\frac{1 + 2\,\tan\theta_{23} + 
    \left( 1 + 8\,\tan^22\theta_{12} \right)
       \,\tan^2\theta_{23}}{8\,
    \tan^22\theta_{12}\,
    \tan^2\theta_{23}}}& \text{for $O_{ij}\ll f_{ij}$}
  \end{cases}\nonumber\\
&\sim
 \begin{cases}
    \displaystyle{\tan^2\theta_{23}
    \left(
      \frac{m_\tau}{m_\mu}
    \right)^2}&\text{for $O_{ij}\gg f_{ij}$}\\
    1& \text{for $O_{ij}\ll f_{ij}$.}
  \end{cases}
\end{align}
As shown in the left part of Fig.~\ref{fig:BR-O33}, in this case
the ratio of branching ratios is larger than one, and therefore an
inverted hierarchy for the leptonic decays of the lightest charged 
Higgs is obtained. In this way, in the small $O_{33}$ case, 
the most important leptonic decay channel 
for the charged Higgs $h^{+}_{1}$ must be 
$h^{+}_{1}\rightarrow(\sum_{i}\nu_{i})\mu^{+}$ instead of 
$h^{+}_{1}\rightarrow(\sum_{i}\nu_{i})\tau^{+}$.

\begin{figure}[t]
  \centering
  \includegraphics[height=6.5cm, width=8cm]{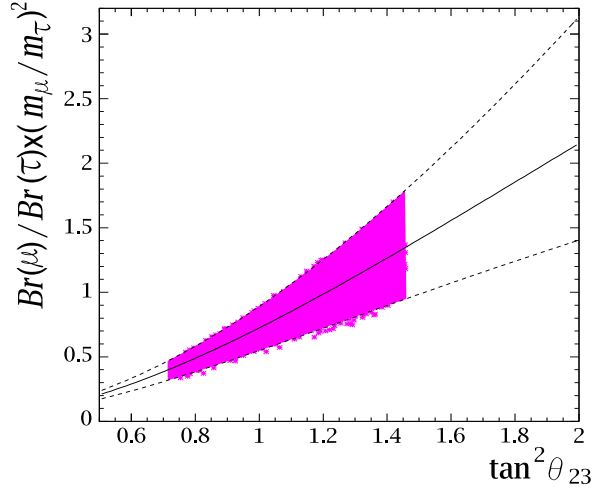}
  \caption{${Br(h_{1}^{+}\to(\sum_{i}\nu_{i})\mu^{+})}/
    {Br(h_{1}^{+}\to(\sum_{i}\nu_{i})\tau^{+})}(m_\mu/m_\tau)^2$ 
    as a function of $\tan\theta_{23}$ in the correlation
    region $O_{ij}\gg f_{ij}$ of the small $O_{33}$ case.
    The solid curve corresponds to the best fit point value of
    $\tan2\theta_{12}$, while the upper and lower curves
    corresponds to its $3\sigma$ limits. See text.}
  \label{fig:correlationhp1}
\end{figure}
Fig.~\ref{fig:correlationhp1} shows the correlation region for
$O_{ij}\gg f_{ij}$ in the small $O_{33}$ case. The curves 
correspond to the ratio ${Br(h_{1}^{+}\to(\sum_{i}\nu_{i})\mu^{+})}/
{Br(h_{1}^{+}\to(\sum_{i}\nu_{i})\tau^{+})}$, normalized by
$(m_\mu/m_\tau)^2$,
as a function of the atmospheric mixing angle, as expected 
from Eq.~(\ref{eq:smallO33}) for the best fit point value 
of $\tan2\theta_{12}$ (solid line) and its 3$\sigma$
limits (dashed lines).
The parameters are fixed as in Fig.~\ref{fig:BR-O33} and the spread of
the points can be understood from the uncertainty in the solar mixing
angle. In this region the charged Higss decay rate 
$\Gamma(h^{+}\rightarrow(\sum_{i}\nu_{i})\mu^{+})$
can be larger than decay rate 
$\Gamma(h^{+}\rightarrow(\sum_{i}\nu_{i})\tau^{+})$ up to a 
factor of $(m_\tau/m_\mu)^2=280$. 

From Fig.~\ref{fig:BR-O33} it can be seen that the large
$O_{33}$ region is divided in three sub-regions, region I
where $10^{-3}
\lesssim Br(h_{1}^{+}\to(\sum_{i}\nu_{i})\mu^{+})/
Br(h_{1}^{+}\to(\sum_{i}\nu_{i})\tau^{+})\lesssim 1$,
region II for which $1
\lesssim Br(h_{1}^{+}\to(\sum_{i}\nu_{i})\mu^{+})/
Br(h_{1}^{+}\to(\sum_{i}\nu_{i})\tau^{+})\lesssim 10^{2}$
and region III characterised by $10^{2}
\lesssim Br(h_{1}^{+}\to(\sum_{i}\nu_{i})\mu^{+})/
Br(h_{1}^{+}\to(\sum_{i}\nu_{i})\tau^{+})\lesssim 10^{4}$.
Measurements of the ratio $Br(h_{1}^{+}\to
(\sum_{i}\nu_{i})\mu^{+})/Br(h_{1}^{+}\to
(\sum_{i}\nu_{i})\tau^{+})$ are sufficient to decide 
whether region I or III are realized.
In region III there is an ambiguity that cannot be removed
by measurements of the ratio $Br(h_{1}^{+}\to
(\sum_{i}\nu_{i})\mu^{+})/Br(h_{1}^{+}\to
(\sum_{i}\nu_{i})\tau^{+})$. However, in the small mixing limit
the ambiguity can be removed. Recalling that in the small
$O_{33}$ region $f_{12}=0$ or $f_{13}=0$ one should expect,
if this region is realized,
\begin{equation}
  \label{eq:cond1}
  Br^{\mu}_s=Br^{e}_s+Br^{\tau}_s\quad\text{or}\quad
  Br^{\tau}_s=Br^{e}_s+Br^{\mu}_s.
\end{equation}
Any deviation from these relations would exclude this region
and in addition with a measurement of the type
$1\lesssim Br(h_{d}^{+}\to(\sum_{i}\nu_{i})\mu^{+})/
Br(h_{d}^{+}\to(\sum_{i}\nu_{i})\tau^{+})\lesssim 10^{2}$
will indicate that region II is realized.
\begin{figure}[t]
  \centering
  \includegraphics[width=8cm, height=6.5cm]{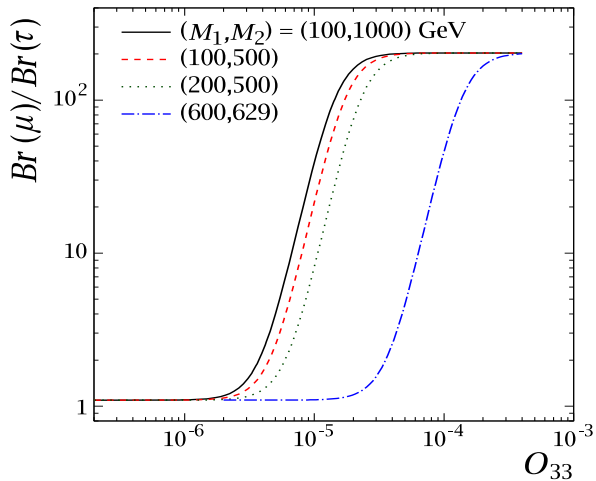}
  \caption{${Br(h_{1}^{+}\to(\sum_{i}\nu_{i})\mu^{+})}/
    {Br(h_{1}^{+}\to(\sum_{i}\nu_{i})\tau^{+})}$ as a function 
    of $O_{33}$ and several pairs of $M_{1}$ and $M_2$ with 
    $\mu=100\,$GeV. From left to right the curves have 
    $\sin2\varphi=0.04$, $0.15$, $0.17$ and $0.99$. The dotted line
    is the same that the curve in the left part of 
    Fig.~\protect{\ref{fig:BR-O33}}. See text.}  
  \label{fig:M_i-dependence}
\end{figure}

The curves in Fig.~\ref{fig:BR-O33} are basically independent of the
value of $\mu$. However, along each curve, smaller values of $O_{33}$ are
excluded as $\mu$ decreases. On the other hand they depend on the
specific value of $M_1$ and $M_2$. In fact, as the mixing angle
$\sin2\varphi$ increases the curves are shifted to the right. This is
illustrated in Fig.~\ref{fig:M_i-dependence} for the small $O_{33}$
case with $f_{12}=0$. All the remaining parameters are chosen as in 
Fig.~\ref{fig:BR-O33}. In particular the dotted line is the same as
the one in Fig.~\ref{fig:BR-O33}.

In summary for the one parameter-solutions we have
\begin{equation}
  \label{eq:summary}
  \left(
    \frac{m_\mu}{m_\tau}
  \right)^2\lesssim  \frac{Br(h_{1}^{+}\to(\sum_{i}\nu_{i})\mu^{+})}
  {Br(h_{1}^{+}\to(\sum_{i}\nu_{i})\tau^{+})}\lesssim10^{4} 
\end{equation}
We have checked that this result holds for all the parameter space of
the NMZM. In particular for $O_{33}$ sufficiently small
the charged Higgs decay rate 
$\Gamma(h^{+}_{1}\rightarrow(\sum_{i}\nu_{i})\mu^{+})$
can be dominant.
\section{Conclusions}
\label{sec:conclusions}
We have considered the version of the Zee model where both Higgs
doublets couple to leptons. Instead of working with 
all the parameters we have focused on a model with
minimal number of couplings consistent with neutrino physics 
data. We have shown that in the small mixing limit 
($\varphi\ll 1$) certain ratios of branching ratios can be 
used to obtain information about the parameters of 
the model. Besides the charged Higgs leptonic decays we have 
also considered the decay $h_{2}^{+}\rightarrow h_{1}^{+}h^{0}$. 
We have found that this decay, if kinematically allowed, can be 
used to determine the value of the $\mu$ parameter. Moreover,
measurements of $\Gamma(h_{2}^{+}\rightarrow h_{1}^{+}h^{0})$ 
allow to decide whether the small mixing limit is realized or not.

Assuming that there are no large hierarchies
among the couplings $O_{i1}$ ($i=1,2,3$) and $O_{ij}$,
and using neutrino physics constraints we have shown
that in this scheme only three parameters are independent.
We have found that there are four regions, in this three-dimensional 
parameter space, determined by only $O_{33}$.
We have shown that two of these four regions are governed
by large values of $O_{33}$ ($10^{-4}-10^{-2}$) 
while the other two regions are governed by small values 
of $O_{33}$ ($10^{-7}-10^{-4}$).

We have analysed charged Higgs leptonic decays in the
large as well as in the small $O_{33}$ 
regimes and we have found: (i) in the large 
$O_{33}$ case, there is a region in which the decays
$h_{1}^{+}\rightarrow \nu_i\mu^{+}$ and
$h_{1}^{+}\rightarrow \nu_i\tau^{+}$
are governed by the correponding Yukawas as in the
2HDM of type-I and type-II and another region
where the decay $h_{1}^{+}\rightarrow \nu_i\mu^{+}$
is enhanced and moreover can be larger than the decay
to $h_{1}^{+}\rightarrow \nu_i\tau^{+}$.
(ii) In the small $O_{33}$ case the decay
$h_{1}^{+}\rightarrow \nu_i\mu^{+}$ is always enhanced
and is larger than the decay
$h_{1}^{+}\rightarrow \nu_i\tau^{+}$. Therefore we suggest
that in order to test the model the decays of the charged 
Higgs to $\nu_i\mu^{+}$ should be searched along with the 
decays to $\nu_i\tau^{+}$. In fact, measurements of 
the ratio of branching ratios
$Br(h_{1}^{+}\to(\sum_{i}\nu_{i})\mu^{+})/
Br(h_{1}^{+}\to(\sum_{i}\nu_{i})\tau^{+})$ could give
information about what region of this parameter space 
is realized.

At future colliders the decay channel $\nu\tau^{+}$
is very important for the discovery of charged Higgs 
bosons \cite{Roy:2005yu,Kanemura:2000cw}. 
For the LHC and SUSY like 2HDM, it has been 
claimed that the existence of a relatively heavy 
charged Higgs bosons, of mass up to 1 TeV, can be 
probed using the signal $h_{1}^{+}\rightarrow \nu\tau^{+}$
\cite{Roy:2005yu}. At future linear colliders a single 
produced charged Higgs should be associated with the 
tau and the neutrino coming from the virtual charged 
Higgs decay \cite{Kanemura:2000cw}.  
According to our results, and illustrated by Fig.~\ref{fig:BR-O33},
the charged Higgs could emerge from a signal with
$\nu_i\mu^{+}$ instead of $\nu_i\tau^{+}$. Moreover, for a 
light charged scalar ($M_{1} < m_{t}$) the ratio 
of branching ratios, $Br(h_{1}^{+}\to(\sum_{i}\nu_{i})\mu^{+})/
Br(h_{1}^{+}\to(\sum_{i}\nu_{i})\tau^{+})$ should be measurable.
\section{Acknowledgments}
We thank W. Porod and E. Nardi for very useful suggestions.
Especially to M. Hirsch for his advice as well as for critical 
readings of the manuscript. D.A. wants to thanks the 
{\it ``Instituto de F\'{\i}sica de la Universidad de 
Antioquia''} for their hospitality. This work was supported
by Spanish grants BFM2002-00345 and FPA2005-01269. 
D.A. is supported by a Spanish PhD fellowship by M.C.Y.T.
D.R. was partially supported by  \textbf{ALFA-EC} funds \footnote{This
document has been produced with the assistance of the European Union.
The contents of this document is the sole responsibility of the authors
and can in no way be taken to reflect the views of the European Union.}.
\appendix
\section{The Three-parameter solution}
\label{apen:oneparamsol}
From the set of Eqs. (\ref{eq:neutrinoMMentries}) 
we choose to express $f_{12}$, $f_{13}$, $f_{23}$ and 
$O_{13}$ in terms of $O_{33}$, $O_{22}$, and $O_{12}$
\begin{align}
  \label{eq:a1}
  f_{12}=&-\frac{A}{B}\frac{m_\tau\,\sqrt{\Delta m_{23}^2}}
  {m_\mu m_\tau}\frac{1}{O_{33}}\\
  \label{eq:a2}
  f_{13}=&
  \left[\frac{ 
      \frac{m_\mu}{m_\tau}\left( \frac{\sqrt{2}\,O_{22}}
        {O_{33}} + \frac{O_{12}\,
          \left( 1 + \tan\theta_{23} \right) }
        {O_{33}\,\tan2\theta_{12}\,
          \tan\theta_{23}} \right)-\sqrt{2}}{\kappa \,
      \,\left( 1 - 
        \frac{m_\mu\,O_{22}}{m_\tau O_{33}} \right)\,\left( 1 + 
        \frac{1}{\tan\theta_{23}} \right) }\right]
  \frac{\sqrt{\Delta m_{23}^2}}{m_\tau}\frac{1}{O_{33}}\\
  \label{eq:a3}
  f_{23}=&\frac{1}
  {\kappa\,\left( 1 - 
      \frac{m_\mu \,O_{22}}{m_\tau\,O_{33}} \right)\,\tan2\theta_{12}}
  \frac{\sqrt{\Delta m_{23}^2}}{m_\tau}\frac{1}{O_{33}}\\
  \label{eq:a4}
  O_{13}=&\frac{ 2\,\sqrt{2}\,O_{12}\,
    \tan2\theta_{12} -
    O_{22}\,\left( 1 + \tan\theta_{23} \right)
  }{2\,\left[ \sqrt{2}\,O_{22}\,
      \tan2\theta_{12}\,
      \tan\theta_{23} + 
      O_{12}\,\left( 1 + \tan\theta_{23} \right)
    \right] }O_{33}
\end{align}
where
\begin{align}
  \label{eq:7}
A=&\left[ 1 + \left( 2 + 
    4\,\tan^22\theta_{12} \right) \,
  \tan\theta_{23} + 
  \tan^2\theta_{23} \right] \nonumber\\
&-2\,\frac{m_\mu}{m_\tau }\,\tan2\theta_{12}\,
\left[ 2\,\frac{O_{22}}{O_{33}}\,\tan2\theta_{12}\,
  \tan\theta_{23} + 
  \sqrt{2}\,\frac{O_{12}}{O_{33}}\,
  \left( 1 + \tan\theta_{23} \right) 
\right]  \\
B=&2\,\kappa \,
    \left( \frac{m_\mu}{m_\tau} \,\frac{O_{22}}{O_{33}} - 
      1 \right) \,
    \tan2\theta_{12}\,
    \left( 1 + \tan\theta_{23} \right) \nonumber\\
   &\times\left[ \sqrt{2}\,\frac{O_{22}}{O_{33}}\,
       \tan2\theta_{12}\,
       \tan\theta_{23} + 
      \frac{O_{12}}{O_{33}}\,\left( 1 + \tan\theta_{23} \right)
          \right]
\end{align}

\end{document}